\documentclass[a4paper,fleqn,usenatbib,useAMS]{mpazh}

\usepackage{graphicx}	
\usepackage{amsmath}	
\usepackage{amssymb}	
\usepackage{multicol}        
\usepackage{bm}		

\usepackage{color}
\usepackage{pdflscape}

\usepackage[T1]{fontenc}
\usepackage{ae,aecompl}
\usepackage{newtxtext,newtxmath}

\newcommand{\ergs}{\,erg\,s$^{-1}$}

\newcommand{\ph}{\,ph\,s$^{-1}$\,cm$^{-2}$}
\newcommand{\ergscm}{\,erg\,s$^{-1}$\,cm$^{-2}$}

\newcommand{\fe}{\ion{Fe}{I}\,K$_{\alpha}$}
\newcommand{\feIb}{\ion{Fe}{I}\,K$_{\beta}$}
\newcommand{\nikel}{\ion{Ni}{I}\,K$_{\alpha}$}

\newcommand{\feK}{\ion{Fe}{XXV}\,K$_{\alpha}$}
\newcommand{\feKb}{\ion{Fe}{XXV}\,K$_{\beta}$}
\newcommand{\feL}{\ion{Fe}{XXVI}\,Ly$_{\alpha}$}
\newcommand{\niK}{\ion{Ni}{XXVII}\,K$_{\alpha}$}
\newcommand{\niL}{\ion{Ni}{XXVIII}\,Ly$_{\alpha}$}

\title[Nickel abundance in SS~433 wind]{An upper limit on nickel overabundance in the supercritical accretion disk wind of SS~433 from X-ray spectroscopy}
\author[Medvedev~et~al.]{Pavel~S.~Medvedev$^{1}$\thanks{E-mail: \href{mailto:tomedvedev@iki.rssi.ru}{tomedvedev@iki.rssi.ru}}, Ildar~I.~Khabibullin$^{2,1}$\thanks{E-mail: \href{mailto:khabibullin@iki.rssi.ru}{khabibullin@iki.rssi.ru}}, Sergey~Yu.~Sazonov$^{1}$, \newauthor
 Eugene~M.~Churazov$^{2,1}$ and Sergey~S.~Tsygankov$^{3,1}$ 
 \\
$^{1}$ Space Research Institute of the Russian Academy of Sciences (IKI), 84/32 Profsoyuznaya Str, Moscow, Russia, 117997 \\
$^{2}$ Max Planck Institute for Astrophysics, Karl-Schwarzschild-Strasse 1, 85741 Garching, Germany\\
$^{3}$ Tuorla Observatory, Department of Physics and Astronomy, University of Turku,
  V\"ais\"al\"antie 20, FI-21500 Piikki\"o, Finland }

\date{Accepted XXX. Received YYY; in original form ZZZ}

\pubyear{2018}

\begin{document}
\label{firstpage}
\pagerange{\pageref{firstpage}--\pageref{lastpage}}
\maketitle

\begin{abstract}
We take advantage of a long (with a total exposure time of 120 ks) X-ray observation of the unique Galactic microquasar SS~433, carried out with the \textit{XMM-Newton} space observatory,  to search for a fluorescent line of neutral (or weakly ionized) nickel at the energy 7.5 keV. We consider two models of the formation of fluorescent lines in the spectrum of SS~433:  1) due to reflection of hard X-ray radiation from a putative central source on the optically thick walls of the accretion disk ``funnel''; and 
2) due to scattering of the radiation coming from the hottest parts of the jets in the optically thin  wind of the system. It is shown, that for these cases,  the photon flux of \nikel\ fluorescent line is expected to be $0.45$ of the flux of \fe\ fluorescent line at 6.4 keV, for the relative nickel overabundance $Z_{{\rm Ni}}/Z = 10$, as observed in the jets of SS~433. 
For the continuum model without the absorption edge of neutral iron, we set a 90 per cent upper limit on the flux of the narrow \nikel\ line at the level of $0.9 \times 10^{-5}$  \ph. For the continuum model with the absorption edge, the corresponding upper limit is  $2.5 \times 10^{-5}$ \ph. At the same time, for the \fe\ line, we measure the flux of $9.9_{8.4}^{11.2} \times 10^{-5}$  \ph. Taken at the face value, the  results imply that the relative overabundance of nickel in the wind of the accretion disc should be at least 1.5 times less than the corresponding excess of nickel observed in the jets of SS~433.
\end{abstract}

\section{Introduction}
\label{sec:intr}
SS 433 is the only Galactic X-ray binary system where accretion of matter onto the compact object permanently proceeds in highly supercritical regime, with the specific accretion rate $\dot{m}=\dot{M}/\dot{M}_{Edd}\sim\,400$, $\dot{M}_{Edd}=3\times 10^{-8} \left(\frac{M}{M_{\odot}}\right)\, M_{\odot}$ yr$^{-1}$, where $M$ is the mass of a compact object (see \citealt{Fabrika2004} for a review). In this case, the standard theory of accretion predicts the presence of intensive matter outflows, both in the form of the accretion disk wind and relativistic jets \citep{SS1973}.  The wind  is expected to start arising from the spherization radius,  $R_{sp}\sim \dot{m}\,R_{in}$, where the inner radius of the accretion disk is $R_{in}=\frac{6GM}{c^2}\,\sim\,10^6\,\left(\frac{M}{M_{\odot}}\right)$ cm,
$G$ is the gravitational constant and $c$ is the speed of light. 
At the same time, the jets are likely to be launched from the region in the very vicinity of the compact object, i.e. at a distance of the order of  $ \sim 10 R_{in}$. 
Although this general picture is confirmed by the recent numerical simulations \citep{Ohsuga2011,Fender2014}, the specific mechanisms that determine actual properties of these outflows, in particular their geometry, velocity and mass loading, remain unclear.  On the other hand, measuring the properties of the outflows from the observational data is of vital importance.

Characteristics of the relativistic jets are relatively well studied thanks to their intense X-ray radiation, that is well described by the model of the nearly ballistic, moderately relativistic matter flow that first becomes visible at a point where its temperature is $T_0\sim 30$ keV\footnote{Hereinafter the temperature is expressed in energy units $kT$, where $k$ is the Boltzmann constant.} (the jet base) and then cools down due to adiabatic expansion and radiative losses until $T\sim 0.1$ keV \citep{Brinkmann1996,Kotani1996,Marshall2002,KMS2016}, where thermal instability develops and causes fragmentation of the flow \citep{Brinkmann1988}. As a result, the X-ray spectrum of jet's radiation is composed by the thermal bremsstrahlung continuum (coming from the hottest parts of the jets) and numerous emission lines from highly-ionized (predominantly H- and He-like) atoms of heavy elements, namely silicon, sulphur, iron and nickel, which are produced mainly in regions with temperatures that ensure the maximum plasma emissivity in a given line.

 Remarkably, the relative intensities of the emission lines, as well as  their ratios to the thermal continuum are broadly consistent with the abundances of heavy elements being close to the solar values (see e.g \citealt{Kotani1996,Marshall2002}). However, the inferred abundance of nickel appears to be $ \sim 10 $ times solar
\citep{Kotani1996,Brinkmann2005,Medvedev2010}. In this case, due to  significant transverse velocity gradient present inside the jets, the observed excess of nickel can not be attributed to e.g. more efficient resonant scattering (and therefore redistribution of the fraction of the photons into the broad line wings due to scattering off hot electrons, \citealt{Sazonov2000}) in the lines 
of more abundant, and therefore having a greater intrinsic optical depth for resonance scattering, elements like iron, sulphur and silicon \citep{KS2012}.
The source of this Ni-enrichment remains unclear, and might be connected both to the site and mechanism of launching the jets, it's further propagation to the place when it becomes directly visible to an observer, and to the peculiarities of the chemical composition of the companion star.

From this point of view, it seems important to try to measure the nickel abundance also in the matter of the wind of the accretion disc, the mass flow in which ($\sim \dot{M} $) in reality can be several orders of magnitude greater than the mass flow in jets (\mbox{$\sim \dot{M}_{Edd}$}, see, e. g., \citealt{KMS2016}). 
The accretion disk wind is likely responsible for obscuration of the hottest parts of the jets,
due to which it manifests itself in the X-ray range by blocking, reprocessing and scattering radiation
from the jets  and possibly also from the putative central source.
The latter scenario was proposed by  \cite{Medvedev2010} to explain the excess of continuous radiation in the SS~433 spectrum above 3 keV with respect to the predictions of the jets emission model, as well as the observed fluorescence line of neutral (or weakly ionized) iron \fe\ with the photon flux at the level of $ 10^{-4}$ \ph\  \citep{Kotani1996,Marshall2002,Brinkmann2005,Kubota2010,Medvedev2010,KMS2016}.

The position (6.4 keV) and the width ($\Delta v_{FWHM}<1000$ km\,s$^{-1}$) of the fluorescent iron line imply that the medium in which it forms is fairly cold and without significant line-of-sight velocity dispersion. However, in order to restore a more complete picture of geometry and physical properties (e.g., the abundance of heavy elements) of this medium, it is also necessary to separate a component directly related to the scattered radiation from the total observed continuum. In principle, such a separation is possible by determining the contribution of the jets emission over the soft X-ray band (below 3 keV),
where numerous emission lines are present.
 In this region, the contribution of the additional hard component is expected to be small and modern X-ray spectrometers (e.g., \textit{Chandra/HETGS}) are characterized by the greatest sensitivity and resolution \citep{KMS2016}. 
Another unique possibility is to study radiation variability, in particular, cross-correlation analysis in different energy ranges.

An additional possibility of studying the environment responsible for fluorescent radiation is associated with the search for fluorescent lines from other elements, which intensity ratio to the intensity of the fluorescent iron line would allow us determining the relative abundance of different elements. Thus, in the case of solar heavy element abundance, the fluorescent nickel line \nikel\ at 7.5 keV  is predicted to be a factor of $20$ weaker than the \fe\ line at 6.4 keV, so that its detection is rather challenging even for sources with sufficiently bright iron line. Nevertheless, the fluorescence nickel line is reported to be detected in the X-ray spectra of obscured active galactic nuclei \citep[AGN,][]{Molendi2003,Fukazawa2016}. However, in the case of a tenfold nickel overabundance in the wind matter, as is observed in the jets of SS~433,  this line becomes only twice weaker than the readily detected \fe\ line, i.e. at the level of $6 \times 10^{-5}$ \ph,  hence it must be feasible to detect it already in currently available data, that amounts to several hundred kiloseconds of SS~433 observations with X-ray spectrometers.

Unfortunately, there are two complications that significantly hinder such an approach. First, the effective area of \textit{Chandra} High Energy Transmission Gratings Spectrometer (\textit{Chandra}/HETGS) drops dramatically above 7 keV, so detection of such a weak line at 7.5 keV is hardly possible, taking into account the spectral variability of the source caused by the ``motion'' of the jet lines, and also because of the absence of the self-consistent continuum model in a given spectral region (e.g., the presence of a neutral iron absorption edge can affect the continuum shape at energies above 7.1 keV).
On the other hand, the \textit{XMM-Newton} EPIC-pn spectrometer  has a substantial effective area up to energy of about 10 keV, which in principle makes it possible to detect the line of interest with a high significance. 
However, here one needs to take into account significantly worse spectral resolution of this instrument, that is at the level of $ \Delta E_{FWHM}=140 $ eV at 7 keV, which potentially leads to contamination of the spectral range of interest by the emission of much brighter and instrumentally-broadened lines of relativistic jets.

In this work we analyze the available \textit{XMM-Newton} archive data and select the observation best suited for searching \nikel\ line, taking into account the sensitivity achieved, the brightness and precessional phase of the source. 
Thus, it is a combination of \textit{XMM-Newton} EPIC-pn high sensitivity and the properly selected observational phase that allows us to constrain Ni abundance in the supercritical accretion disc wind using X-ray fluorescence for the first time.

The paper is organised as follows: in Section~\ref{sec:data}, we
describe the data, and put it in the context of the system's precession variability and previous observations. In Section~\ref{sec:models}, we describe models of the spectral components that contribute to the observed emission in the explored spectral domain. In Section~\ref{sec:results}, we explore the data in the light of the spectral components described before, and end up with the measurements and constraints on the line fluxes of \fe\ and \nikel. We discuss implications of the obtained estimate of the \nikel-to-\fe\ ratio for the inferred nickel-to-iron abundance ratio in Section~\ref{sec:discussion}. Summary of the paper follows in Section~\ref{sec:conclusions}. A detailed description of the method of estimating the confidence regions, which are of vital importance for understanding the compatibility of nickel abundance in the jets and in the wind of the system, is given in Appendix~\ref{sec:append}.

\section{Data}
\label{sec:data}
\begin{figure}
\centering
\includegraphics[width=0.45\textwidth,bb=50 180 550 670]{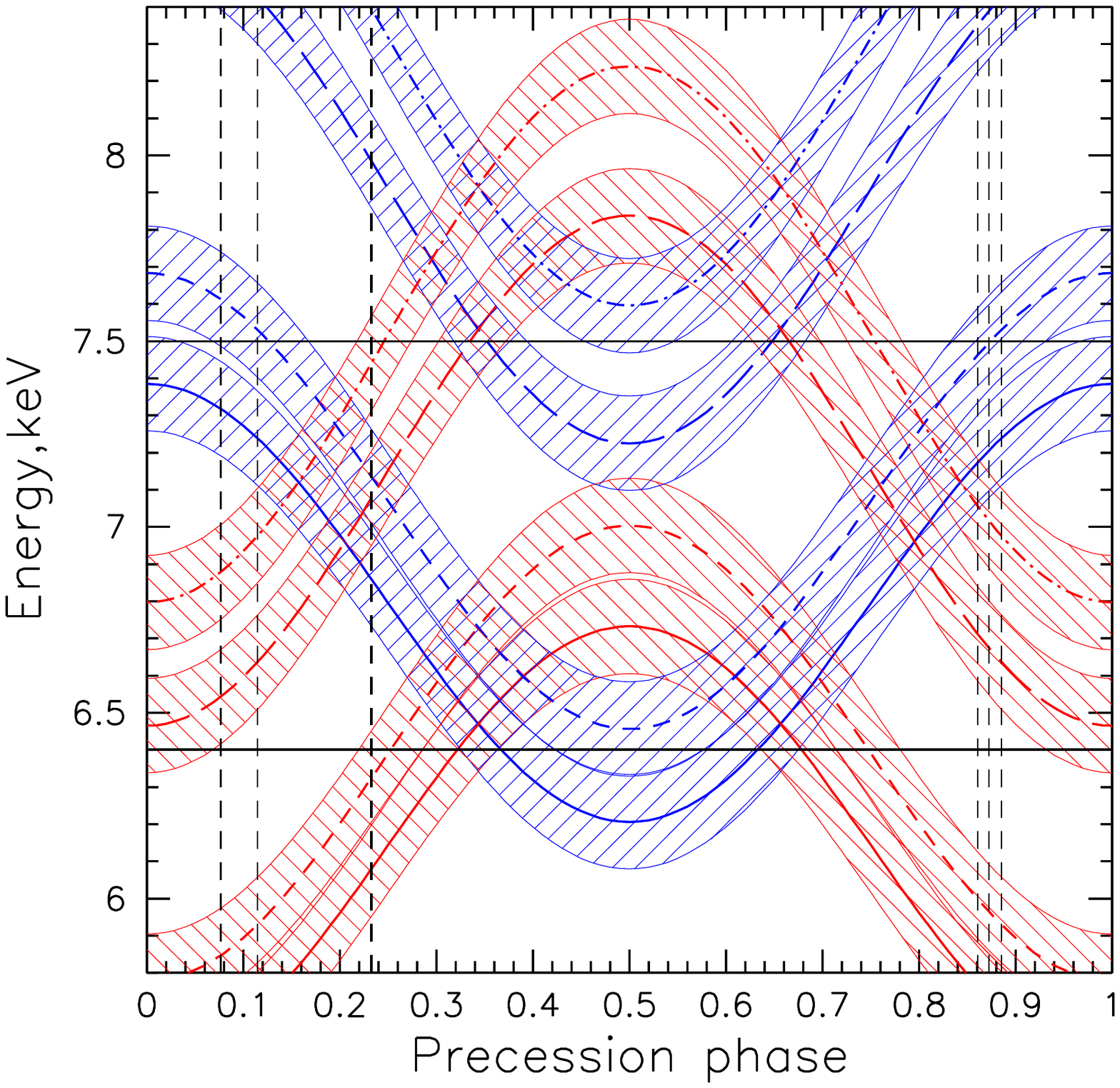}
\caption{\small 
Predicted positions of the brightest X-ray lines of the approaching (blue lines) and of the receding (red lines) from the observer jets
are shown as a function of precession phase of SS~433.
 The lines are \feK\ ($E_0=6.7$ keV, solid lines), \feL\ ($E_0=6.96$ keV, short dashed), \niK\ ($E_0=7.8$ keV, long dashed) and \niL\ ($E_0=8.1$  keV, dash-dotted line). The hatched areas depict the ``region of influence'' for each line. This region is defined so that the intensity of the line emission convolved with  the \textit{XMM-Newton} EPIC-pn  spectral response function drops
  10 times compared to the peak value at the region boundary (i.e., the half-width $ \delta E=\sqrt{2\ln(10)}\Sigma_{E}$ and $\Sigma_E\approx 60$ eV correspond to $\Delta E_{FWHM}=140$ eV). It is assumed that the positions of the \fe\ (6.4 keV) and \nikel\ (7.5 keV) fluorescent lines 
  do not change with precession phase; they are shown by horizontal black solid lines. The precession phases of the available \textit{XMM-Newton} non-eclipse observations are depicted with vertical dashed lines, with the thick line corresponding to the observation analysed here.}
\label{f:eefluor}
\end{figure}

We survey all non-eclipse observations of SS~433 in the \textit{XMM-Newton} data archive to select the observations best suitable for searching for the fluorescent line \nikel\ at 7.5 keV in terms of ``contamination'' of the region of interest by the much brighter emission lines of the jets. The results of this analysis are shown in Fig.~\ref{f:eefluor}, where the vertical dashed lines show all non-eclipse data. Observation of 2012 (\textit{ObsID} 694870201, PI: Aleksei Medvedev), which served as the basis for the current work (shown by thick dashed line), corresponds to the precessional and orbital phases of SS~433 $\psi=0.24$ and $\phi=0.44$--0.55, respectively \citep[according to ephemeris by][]{Goranskij2011}. The spectral lines of the approaching jet are depicted with blue curves in Fig.~\ref{f:eefluor}. These are the brightest lines in the X-ray spectrum of SS~433  and are typically 3--4 times brighter than the corresponding lines of the receding jet (marked by red curves), as a result of the relativistic boosting and possible additional absorption of the light from the receding jet 
(see \citealt{Kotani1996,Marshall2002} for details). For this reason, the considered observation is indeed the most optimal of all available in the \textit{XMM-Newton} archival data.


The \textit{XMM-Newton} observation \textit{ObsID} 694870201 was performed on October 3--5, 2012. Data reduction was carried out using the standard science analysis package (SAS) version 13.0. As the first step, we have selected all events, that were not affected by proton flares. The net effective exposure for the EPIC-pn instrument was $\approx 124$ ks.

\begin{figure}
\centering
\includegraphics[width=1\columnwidth,bb=50 25 600 552]{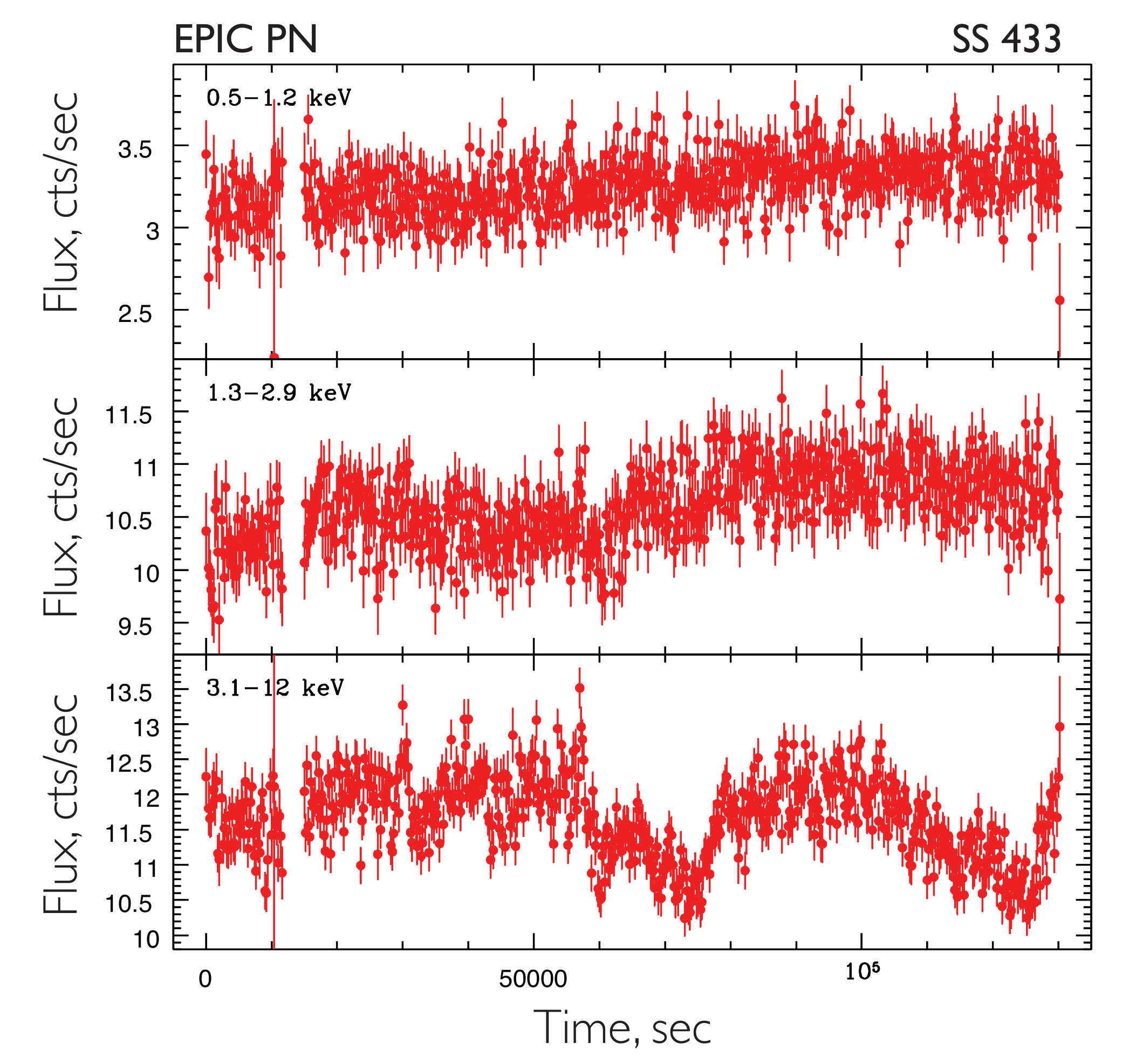}
\caption{\small
EPIC-pn light curves of SS~433 during the \textit{XMM-Newton} observation on
 October 3--5, 2012 in the 0.5--1.2 keV (top panel), 1.3-2.9 keV (middle panel), 3.1-12 keV (bottom panel) energy bands.} 
\label{f:light_curve}
\end{figure}

The EPIC-pn camera was operated in the timing mode suited for observations of bright sources.
The average count rate from SS~433 was about 25 cts\,s$^{-1}$.
The final processing of the data was carried out in accordance with the standard user manual\footnote{\url{www.cosmos.esa.int/web/xmm-newton/sas-threads/}}. 
In particular, background photons were extracted from rows of pixels 2--16, and photons related to the source were taken from rows 29--47.  As a result, we obtain light curves shown in Fig.~\ref{f:light_curve}.  Finally, the data were averaged for both the full exposure and for 10 ks, allowing the spectral evolution of the source to be studied with sufficient statistics. 
In this work, we use spectra without binning the energy channels and apply the weighting function by \cite{Churazov1996} instead (unbinned weighted spectra). Besides that, we have checked that analysis of the spectra, re-binned with at least 25 raw counts per bin by means of the standard tool \texttt{GRPPHA}, gives the results that are consistent within the 90 \% confidence ranges (see also Appendix~\ref{sec:append}).

\section{Spectral models} 
\label{sec:models}

\subsection{Baryonic jet} 
\label{sec:baryonicjet}

The spectral model of the thermal X-ray emission from a baryonic jet with the parameters relevant for SS~433 (hereafter \texttt{bjet} model) has been calculated and made publicly available by \cite{KMS2016}. 
This model is based on the solution of the thermal balance equation with the radiative losses taken into account self-consistently, what allows very accurate reconstruction of the emission measure distribution over the range of temperatures responsible for continuum (mostly the hottest parts of the jet) and line (part of the jet with $ T<10 $ keV) emission. Total radiation of the jet is then found by summing up contributions from thin isothermal transverse slices along the jet, that are calculated in optically thin collisional ionization equilibrium (CIE) model, based on the \texttt{AtomDB/APEC} atomic database\footnote{\url{http://www.atomdb.org}} \citep[version 3.0.9, ][]{Foster2012}.
The set of input parameters for the SS~433-adjusted version of the \texttt{bjet} model (with fixed the bulk velocity of the jet matter, $\beta=v/c=0.26$ and the half-opening angle, $\Theta = 0.02$ rad) includes the jet kinetic luminosity $L_k$, the gas temperature at the jet base $T_0$ (the jet base is the directly observable jet region closest to the compact object), 
the transversal optical depth at the jet base with respect to electron scattering
$\tau_{e0}$, the abundance of elements heavier than helium $Z$ (with the exception of nickel), and also 
nickel abundance $Z_{{\rm Ni}}$, allowed to vary separately from other heavy elements. The abundance parameters are quantified relative to the set of solar abundance of \cite{Anders1989}. The model predicts both the shape of the emitted spectrum and the total X-ray luminosity of the jet.
The spectral shape is determined mainly by the shape of the differential emission measure distribution along the jet and 
actually depends almost solely on the value $\alpha$, which is a combination of model parameters:
\begin{equation}
\alpha = \frac{2}{3}\frac{\tau_{e0}}{\Theta\beta}\frac{\Lambda_Z(T_0)}{\sigma_e c T_0} \frac{X}{1+X},
\end{equation}
where $\Lambda_Z(T_0)$ is the integrated plasma emissivity, $X = n_i/n_e \approx 0.91$ is the ion-to-electron ratio, $\sigma_e=6.65 \cdot 10^{-25}$ cm$^2$ is the Thomson cross-section.  Assuming that radiative cooling is determined only by hydrogen and helium bremsstrahlung, one can get a simple estimate: $\alpha \approx  4.42~\tau_{e0} \times~\left(\frac{10~{\rm keV}}{T_{0}}\right)^{1/2}$. The physical meaning of the parameter $\alpha$ is the ratio of the radiative cooling term to the adiabatic one in the thermal balance equation at the jet's base. 
Since for  $ \alpha \ll 1$  this ratio remains small across almost the whole jet, the cooling is determined by adiabatic expansion in this case.
For the $\alpha\gg 1$, the cooling is dominated by radiative losses
\citep[for more details, see ][]{KMS2016}.

\begin{figure}
\centering
\includegraphics[width=1\columnwidth]{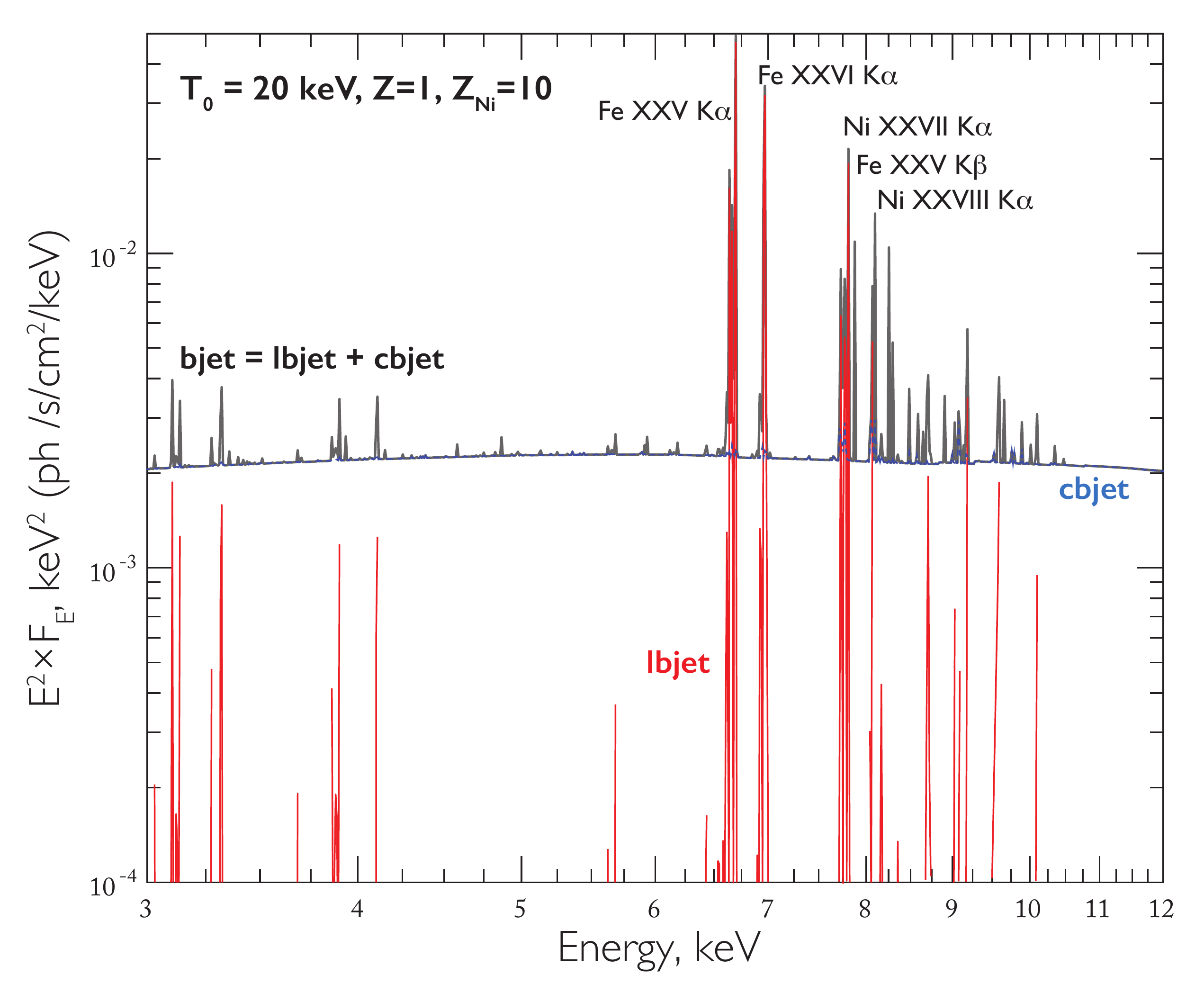}
\vspace{-10pt}
\caption{\small
The spectral models of thermal X-ray emission from the baryonic jets of the SS~433 system. The black line shows the spectrum obtained in the \texttt{bjet} model for the jet parameters $L_k\times \tau_{e0} = 0.5 \times 10^{38}$ \ergs, $T_0 = 20$ keV, $Z = 1$, $Z_{{\rm Ni}} = 10$. The blue dotted line shows the \texttt{cbjet} model, which is the continuum component  (in sum with the pseudo-continuum) of the total predicted spectrum. The red line shows the \texttt{lbjet} line emission model used in the paper, determined by the difference \texttt{bjet}$-$\texttt{cbjet}. The predicted flux is reduced to a distance $D=10$ kpc to the source. 
We marked with the symbols the jet lines that are considered within the framework of the phenomenological model in Section~\ref{sec:gauss}.}
\label{f:lbjet}
\end{figure}

Besides that, one can use individual sub-components of the \texttt{APEC} model, i.e. specific line and pseudo-continuum\footnote{In addition to thermal bremsstrahlung continuum, \texttt{APEC} pseudo-continuum also includes two-photon and recombination continua and also some very weak lines} emissivities at a given temperature, to calculate corresponding components in the integrated jet's spectrum. 
In this work, we use this possibility to extract from the total emergent spectrum a separate component of the continuum emission.
As a result, we calculated a ``continuum''-version of the model presented in \cite{KMS2016} (hereafter \texttt{cbjet} model\footnote{The model is available for downloading from the web-site: \url{http://hea133.iki.rssi.ru/public/bjet/}}), that is calculated on the same grid of parameters and is defined in identical energy bins. Then, subtraction of this model from the full one provides us with the ``lines-only'' spectral model (hereafter \texttt{lbjet=bjet-cbjet} model). We take advantage of the \texttt{lbjet} model to fit the lines observed in the region of interest here.
Typical simulated emergent spectra in the  \texttt{bjet}, \texttt{cbjet}, and \texttt{lbjet} models 
for the same set of parameters are shown in Fig.~\ref{f:lbjet}.

In Section~\ref{sec:results}, in addition to analyzing data using the \texttt{lbjet} model, we also explore the phenomenological approach (hereafter referred to as the phenomenological model) to describe the main spectral features in the 6--9 keV energy range in a form of individual Gaussian lines. \cite{KMS2016}  calculated model predictions for a similar set of basic observables, that can be readily derived even from a low-resolution spectrum without a sophisticated full model fitting involved. 
These observables include total flux and spectral slope of the emission in the line-free band from 3 to 6 keV, fluxes in softer (1.5--3 keV) and harder (6--9 keV) bands and photon fluxes of the brightest lines and their ratios. One can use them to get an idea regarding the temperature at the jet base, abundance of heavy elements and also the presence of the additional hard component in the spectrum (see \citealt{KMS2016} for an example of such a technique tested on \textit{Chandra}/HETGS data). 
The agreement in the predictions of the jets parameters between the two approaches should serve as an additional criterion for the adequacy of the used models.

\subsection{Optically thick reflection} 
\label{sec:reflection}
\begin{figure}
\centering
\includegraphics[width=0.45\textwidth,bb=70 200 550 650]{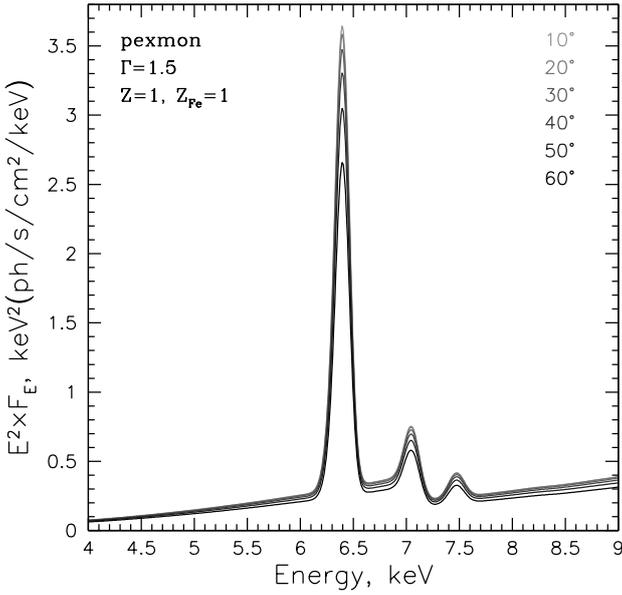}
\caption{\small  
Predicted shapes of the reflected spectrum in the \texttt{pexmon} model  (with no direct component from illuminating source)
are shown for different values of the angle $i$ between the line of sight and the normal to the semi-infinite slab of neutral matter, illuminated by isotropic X-ray source.
The spectrum of the illuminating radiation is a power-law  with photon index $\Gamma=1.5$. The abundance of heavy elements in the reflecting matter is set equal to the solar value.}
\label{f:pexmon}
\end{figure}
\cite{Medvedev2010} put forward a model that attributes hard component and the \fe\ fluorescent line to the reflection of radiation from the putative central X-ray source on the walls of the accretion disk ``funnel''. The luminosity of the central source in this model needs to be very high, at the level of $\sim 10^{40}$ \ergs, what, however, may indeed be true, bearing in mind the highly supercritical regime of accretion taking place in the system and the fact that we observe the system nearly edge-on, so the collimated emission can be easily hidden by the thick supercritical accretion disk. Were SS~433 observed face-on, it would then appear as an ultraluminous X-ray source (ULX, \citealt{Begelman2006,Poutanen2007}; but see \citealt{KS2016} for an upper limit on luminosity).

As an approximation to the spectrum of reflection component arising in such a scenario, we use the \texttt{pexmon} model (\citealt{Nandra2007}, see also paper by \citealt{George1991}, on which it is based)
available in {\sc XSPEC} \citep{Arnaud1996}, that self-consistently reproduce iron and nickel fluorescent lines, along with the scattered continuum and absorption edges corresponding to these elements.  
The geometry of the problem assumed in the \texttt{pexmon} model implies an isotropic illumination of a semi-infinite uniform slab of neutral matter by a source with a power law spectrum of photon index $\Gamma $ and  high energy exponential cut-off proportional to $e^{-E/E_c}$. For fixed $\Gamma$ and $E_c$ the shape of the reflected spectrum is determined by the inclination angle $i$,
defined as the angle between the normal to the slab and the line of sight. In the scenario proposed by \cite{Medvedev2010}, the line-of-sight inclination angle  with respect to the walls of the accretion disc ``funnel'' is expected to be close to perpendicular. 
In Fig.~\ref{f:pexmon} we show how the reflected spectrum changes as the angle $i$ varies from $10^\circ$ to $60^\circ$.
The spectrum of the source is assumed to be a power law with photon index of $\Gamma=1.5$, which roughly corresponds to the spectrum of the hot parts of the SS~433 jets. In addition to the \fe\ line, the fluorescence iron line \feIb\ at 7.06 keV, emitted upon the transition of M-shell electrons, is clearly visible. The flux of the \feIb\ line is $\approx 11 \%$ of the \fe\ line flux \citep[e.g.,][]{Kaastra1993}.

\begin{figure}
\centering
\includegraphics[width=0.45\textwidth,bb=70 200 550 650]{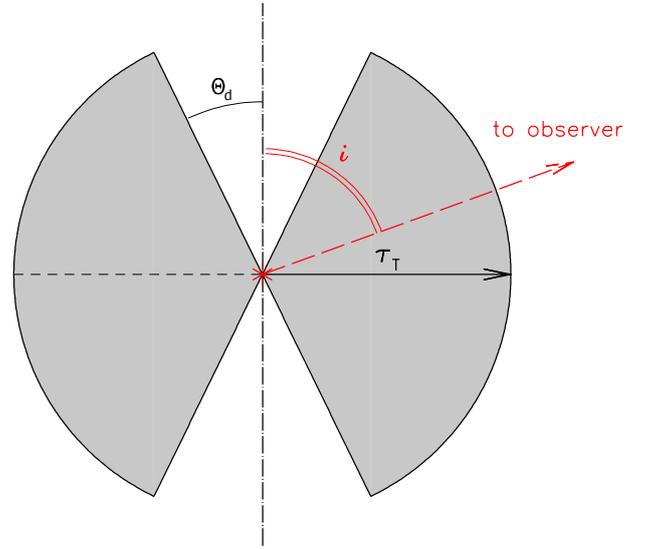}
\caption{\small 
Assumed geometry in the \texttt{cwind} model for optically thin scattering. 
A point illuminating source is considered to be placed in the center of a spherical homogeneous cloud of neutral gas with  two symmetrical conical funnels with half-opening angle $\Theta_d = \arccos\mu_d$ 
excised along the axis passing through the center of cloud. The cloud is characterized by Thomson optical depth in radial direction $ \tau_{T} $, while the direction to the observer is set by the inclination angle $ i=\arccos\mu $ with respect to the aforementioned axis.}
\label{f:sketch_wind}
\end{figure}
\begin{figure*}
\centering
\includegraphics[width=0.8\textwidth]{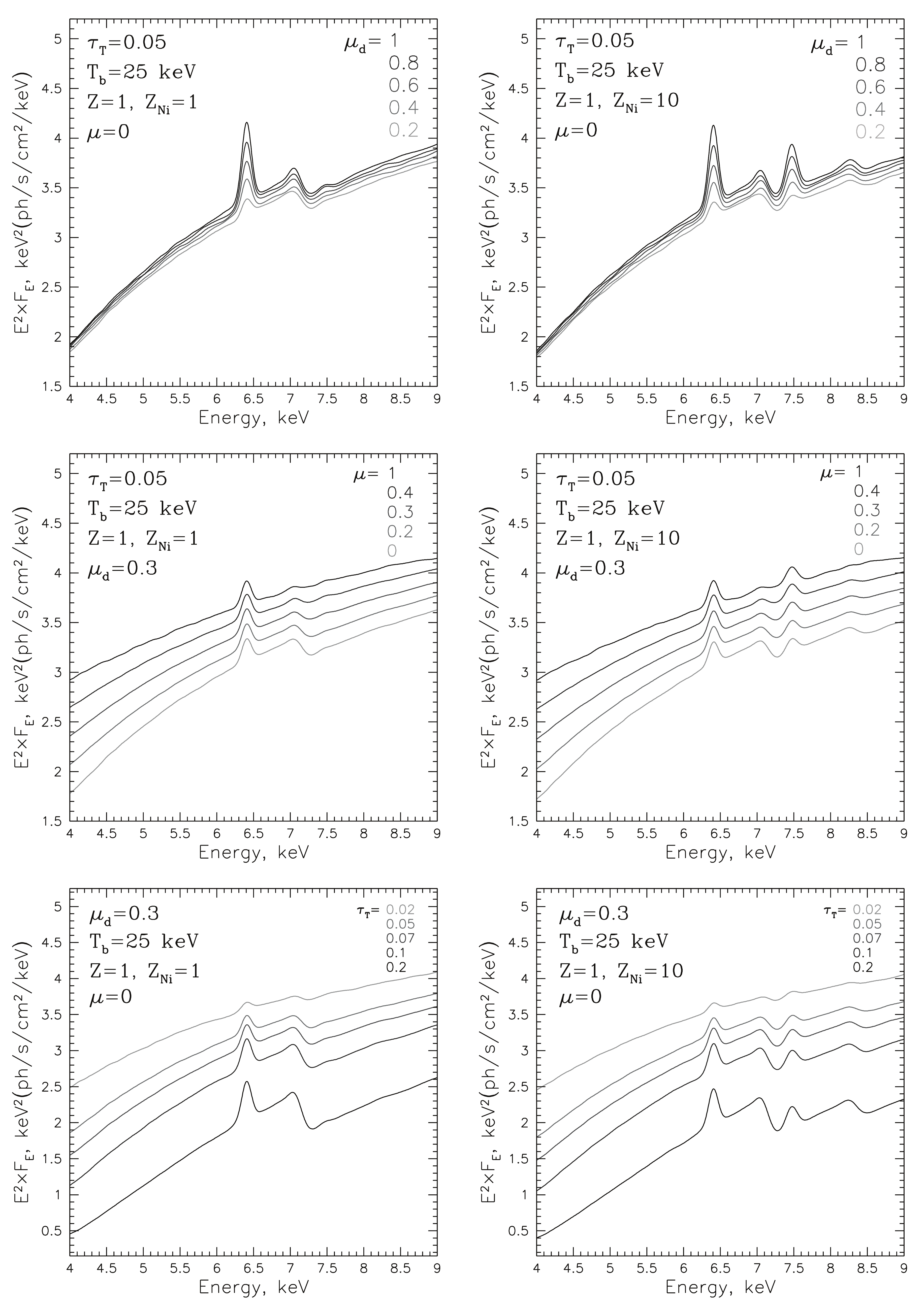}
\caption{\small
X-ray emission from a source with thermal bremsstrahlung spectrum of $T_b=25$ keV scattered by an optically thin cloud of cold gas calculated using the \texttt{cwind} model for different sets of model parameters. The left-hand panels show the results for the solar chemical composition of the obscuring matter, the right-hand panels --- for the relative abundance of nickel $Z_{{\rm Ni}}/Z$ = 10. The two top panels show the spectrum of the scattered component for different values of the  half-opening angle of the
conical wind funnel of the supercritical disc,  $\mu_d$ (see Fig.~\ref{f:sketch_wind}). The two middle panels show the dependence on the viewing angle $\mu = \cos{i}$. Spectra for different values of the radial Thomson optical depth of the cloud $ \tau_{T}$  are shown in the bottom panels.}
\label{f:cwind_all}
\end{figure*}

Analysis of the spectra predicted by the model shows a very weak dependence of the \nikel-to-\fe\ flux ratio on 
the viewing angle (angle $i$).
The fluorescent yield of K-shells increases with the nuclear charge (see \citealt{Bambynek1972} for a review). For the solar element abundances by \cite{Anders1989}, the photon flux ratio of the fluorescent lines of nickel and iron can be roughly estimated as
\begin{equation}
 R_{{\rm fluor}}(Z_{{\rm Ni}}=1) = 
\frac{F({\rm Ni\,I\ K}_\alpha)}{F({\rm Fe\,I\ K}_\alpha)} \sim  \frac{n_{{\rm Ni I}}}{n_{{\rm Fe I}}} \frac{\omega_{{\rm Ni\,K\,\alpha}}}{\omega_{{\rm Fe\,K\,\alpha}}} \approx  0.045,
\end{equation}
where $\omega_{{\rm Ni\,K\,\alpha}}=0.41$ and $\omega_{{\rm Fe\,K\,\alpha}}=0.34$ are the fluorescent yields for \nikel\ and \fe, respectively \citep{Bambynek1972}. 
Such a simple estimate is in good agreement with the predicted $R_{{\rm fluor}}$ value in \texttt{pexmon} model assuming solar element abundance. As long as the probability of photoabsorption of the incident photon by nickel atoms is less than the total probability 
of being absorbed by other elements or scattered by an electron, the flux in the \nikel\ line, and therefore the ratio $R_{{\rm fluor}}$, increases almost linearly with increasing nickel abundance in the reflecting medium.
In the region of interest, $Z_{{\rm Ni}} \sim 10$, this is indeed the case, because the 
opacity of reflecting medium turns out to be dominated by nickel atoms photoabsorption only for $Z_{{\rm Ni}}>20$. 
Therefore, in the case of a tenfold nickel overabundace in the wind of SS~433 supercritical accretion disc, the flux ratio is expected to be at the level of $R_{{\rm fluor}}(Z_{{\rm Ni}}=10) \approx 0.45$. The \texttt{pexmon} model does not allow verifying this prediction, but we test this statement using the \texttt{cwind} scattering model in the next subsection (see Fig.~\ref{f:niferat}). We note that our conclusions based on the \texttt{pexmon} model are consistent with the results of observations and modeling of the reflected component in the spectra of highly obscured AGN \citep{Yaqoob2011,Molendi2003}.


\subsection{Optically thin scattering} 
\label{sec:scattering}
In this paper, we for the first time propose an alternative explanation for the appearance of the hard component and the fluorescent iron line in the X-ray spectrum of SS~433, namely that the hard component 
arises from emission of the hottest parts of the jets that are seen through the accretion disk wind, 
which is optically thin with respect to electron scattering (with $\tau_T\lesssim 0.1$), but is optically thick for the photoabsorption at energies below 3 keV.  
In this model, no bright central source needs to be invoked since emission above 4 keV is not dramatically attenuated, although this component does not make any contribution below 3 keV due to steep increase of the photoabsorption cross-section. The fluorescent line of iron arises along with the Thomson scattered continuum, contribution of which to the total observed spectrum turns out, however, to be at the $\tau_T$ level, i.e. is relatively small.

\begin{figure}
\centering
\includegraphics[width=1.1\columnwidth,bb=20 30 620 550]{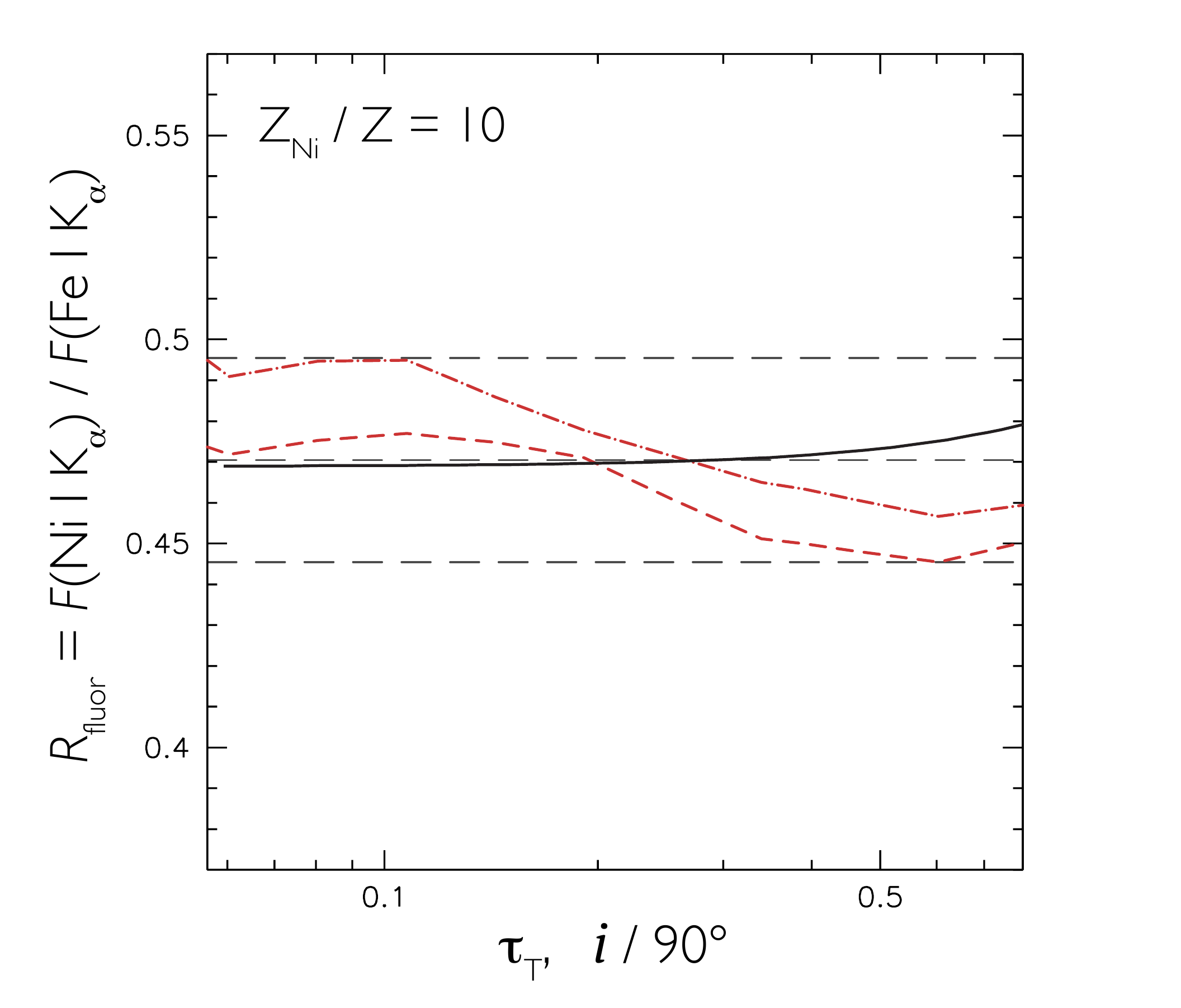}
\caption{\small 
Predicted photon flux ratio of the fluorescent nickel line \nikel\ to the \fe\ line for the two models of their formation: 
a) as the reflection of X-ray radiation from an optically thick neutral medium (\texttt{pexmon} model, solid line), in this case, 
as a function of the inclination angle $i$ between the line of sight and the normal to the reflecting slab; 
2) as the scattering of radiation by an optically thin cloud, calculated in the \texttt{cwind} model for 
two values of the bremsstrahlung temperature: $T_0=25$ keV (red dashed line) and $T_0=35$ (red dotted line). 
In this case, the ratio is shown as a function of the radial Thomson optical depth of the cloud.
The flux ratio is shown for the relative nickel abundance in the reflecting/scattering matter $Z_{{\rm Ni}}/Z = 10$
in solar abundance units (with $Z=1$). The black dashed horizontal lines depict the minimum, average, and maximum values of $R_{{\rm fluor}}$.}
\label{f:niferat}
\end{figure}

We perform Monte-Carlo simulations of the emergent spectrum using the spectral model \texttt{cwind}. A detailed description of the code is given in \cite{Churazov2017b}, where a similar model,
aimed at prediction of X-ray emission reflected by a molecular cloud, is presented
(\texttt{crefl} model\footnote{\url{http://www.mpa-garching.mpg.de/~churazov/crefl/}}). 
Additionally,  a similar model for AGN was presented in \cite{Sazonov2015}. 
In a nutshell, the model accounts for elastic and inelastic scattering, photo absorption of X-ray photons 
and fluorescence on neutral atoms of the most abundant heavy elements.


Geometry of the problem, assumed in the \texttt{cwind} model, is depicted in Fig.~\ref{f:sketch_wind}. Essentially, it implies a point source located in the center of a uniform spherical cloud of neutral gas with 
Thomson optical depth in radial direction $ \tau_{T} $ and abundance of heavy elements $Z$,
in which two symmetrical conical funnels with half-opening angle $\Theta_d = \arccos\mu_d$ are excised along 
the axis passing through the center of cloud and forming the angle $ i=\arccos\mu $ with the observer's line-of-sight.  The source spectrum is given by thermal bremsstrahlung emission with a given temperature $T_b$.
In addition, the relative nickel abundance $Z_{{\rm Ni}}$ in solar units is included as an additional parameter of the model.

Despite a drastically simplified geometry of the problem in the  \texttt{cwind} model, it allows us to cover a wide variety of spectral forms that arise as a result of combining scattered and directly transmitted radiation, for various values of the parameters   $\tau_T$, $\mu$, $\mu_d$ and $Z_{{\rm Ni}}$ (see Fig.~\ref{f:cwind_all}). In the same time, for a reasonable range of gas temperatures at the jet base ($T_b\sim20$--$30$ keV), the spectral shape of the radiation varies only slightly with variations in the parameter $T_b$.

The photon flux ratio of the fluorescent lines $R_{{\rm fluor}} = F({\rm Ni\,I\ K}_\alpha) / F({\rm Fe\,I\ K}_\alpha)$  calculated in the \texttt{cwind} model is shown in Fig.~\ref{f:niferat} as a function of the radial Thomson optical depth $\tau_T$ for the relative nickel abundance $Z_{{\rm Ni}}/Z=10$ ($Z=1$) and for two values of $T_b$: 25 keV (red dashed line) and 35 keV (red dash-dotted line).
In the same figure, the black solid line shows the value of $R_{{\rm fluor}}$  obtained with in \texttt{pexmon} model and {\it multiplied by 10}, depending on the viewing angle $i$, defined as the angle between the line of sight and the normal to the reflecting slab.
As can be seen from the figure, the flux ratio lies in the range 0.45--0.5 and depends only slightly on the model parameters. The linear dependence of the value of $R_{{\rm fluor}}$ on the relative nickel abundance is well confirmed by the agreement of the results of the two models.


\section{Results} 
\label{sec:results}
\begin{figure*}
\centering
\includegraphics[width=1\textwidth]{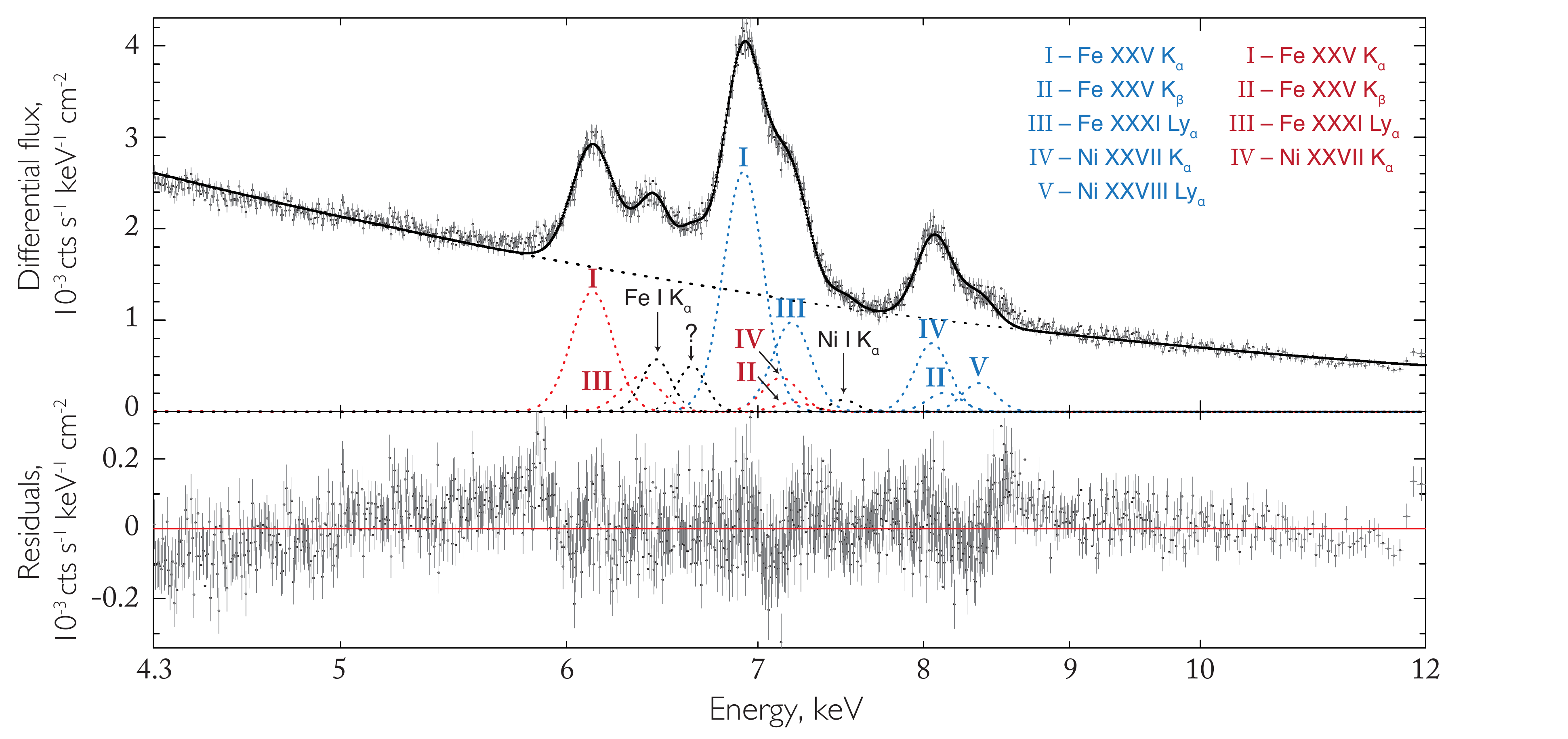}
\vspace{-10pt}
\caption{\small
Phenomenological description of the jets emission lines along with the \fe\ and \nikel\ fluorescent lines in the SS~433 spectrum obtained with the \textit{XMM-Newton} EPIC-pn camera with 120 ks exposure. Blue and red signatures show the lines of the approaching and receding jet, respectively. Black signatures show fluorescent lines, as well as an unidentified spectral line, which 
should be added in order to get a satisfactory fit to the observed spectrum (depicted with a question mark).}
\label{f:bremss_gauss}
\end{figure*}
\subsection{Continuum model}
\label{sec:bremss}
As a first step of working with data, it is necessary to determine the shape and level of the continuum in the 6--9 keV energy range, where, in addition to the fluorescent lines of interest, the brightest emission lines of the jets are emitted. Unfortunately, it is not possible to do this completely self-consistently, since there is still no clear understanding of the processes responsible for the formation of continuous radiation in this spectral region. As was shown by \cite{KMS2016}, the \texttt{bjet} spectral model can well describe the SS~433 spectrum in the soft X-ray band ($<3$ keV) even in the case of data with a high spectral resolution. 
However, at higher energies, an appreciable excess of hard radiation is observed. In particular, this is manifested by the slope of the spectrum in the 3--6 keV range, which does not contain strong spectral lines.
The observed slope is too flat (the photon index $\Gamma\lesssim1.5$) and can not be reproduced by the multitemperature jet model alone
 (see Fig.~7 in \citealt{KMS2016}, as well as \citealt{Brinkmann2005,Medvedev2010}). It is quite possible that the contribution 
 to the hard energy band is made by emission from the hottest parts of the jets scattered by the optically thin wind and
  being suppressed in the soft band due to photoabsorption (see Section~\ref{sec:models}). 
 In these terms it is worth noting the study of SS~433 on a large volume of \textit{RXTE} data \citep{Filippova2006}, which showed that the SS~433 spectrum at energies up to 50 keV is in good agreement with a thermal bremsstrahlung emission with temperature of $\sim 20$--$30$ keV. 
 As has been noted in \cite{Marshall2002}, there is the interesting agreement between 
 the sound speed of the flowing gas at the aforementioned temperature and the transverse expansion velocity of the jet with the opening
angle of $\sim 1.5$ degrees.  
Nevertheless, other scenarios for the formation of the hard component are possible, such as the reflection model \citep{Medvedev2010}, or  Comptonization of soft X-ray photons on hot electrons of the hypothetical corona of the accretion disc \citep{Cherepashchuk2009,Krivosheyev2009}.

We are interested in the search for spectral features near the \nikel\ line at 7.5 keV. 
Therefore, for the current study, we simplify the continuum to a single-temperature thermal bremsstrahlung fit, 
independent of the line components.  To this end, we derive the continuum parameters in the spectral regions that do not contain bright emission lines: 4.3--5.8 keV and 10--12 keV\footnote{For the 4.3--5.8 keV range the lower limit is determined so as to exclude the doublet \ion{Ca}{XX}\,Ly$_\alpha$ 4.1 keV of the approaching jet, and the upper limit excludes the triplet of  \ion{Fe}{XXV}\,K$_\alpha$ 6.7 keV of the receding jet. The lower limit of the 10--12 keV  range excludes the line \ion{Ni}{XXVIII}\,Ly$_\beta$ 9.6 keV of the approaching jet, while the EPIC-pn effective area is almost equal to zero for energies above 12 keV.}. 
The obtained  temperature of $T_{{\rm bremss}} = 22.3_{21.5}^{23.2}$ keV and the flux in 6--9 keV band of $F_{6-9} =  4.22_{4.20}^{4.24} \times 10^{-11}$   \ergscm\ are fixed for further analysis of the spectral lines. 
Here and below, the range of the parameter values is indicated for 90\% confidence level (we describe the method of estimating the confidence regions in Appendix~\ref{sec:append}). The relatively low quality of the fit, corresponding to $\chi^2/{\rm d.o.f} = 985 / 696 = 1.4$, is related to the spectral variability of the source on the time scale of total exposure time
(see Section~\ref{sec:10ks}).

\subsection{Phenomenological method of line description}
\label{sec:gauss}

We begin our analysis of the observed spectrum using simple spectral characteristics \citep[see also][]{KMS2016}, describing strong emission lines in the 6--9 keV band in a form of individual Gaussian lines, linked with the Doppler shifts corresponding to  
the radial velocities of the jets. We will call such a method as a phenomenological model. The brightest observed X-ray lines in the energy range of interest correspond to the K$_{\alpha}$ transitions in helium-like and to the Ly$_{\alpha}$ transitions in hydrogen-like ions of iron and nickel. In addition, we consider the line emitted by the  K$_{\beta}$ transitions in helium-like iron ions, since it falls in the vicinity of the \niK\ and \niL\ lines that are of importance for the current study. We call the jet approaching to the observer as the ``blue jet'' (lower index ``b''), and the oppositely directed jet as the ``red jet'' (lower index ``r'').

The \fe\ fluorescent iron line  is described by a narrow Gaussian with a width of $\Sigma_{{\rm Fe\,I}} = 5$ eV\footnote{The width of a line with a Gaussian profile will be called the dispersion parameter $\Sigma_{E0}$ for the standard form of the Gauss function: $I(E)= F \frac{1}{(1+z) \Sigma_{E0} \sqrt{2\pi}} exp{\frac{-(E(1+z)-E_0)^2}{2 \Sigma_{E0}^2}}$, where $z$ is the Doppler shift, $F$ and $E_0$ --- flux and centroid of a line in the the rest frame of the emitter.} and a centroid energy at $E_{{\rm Fe\,I}}=6.4$ keV (as inferred from \textit{Chandra}/HETGS data, \citealt{Marshall2002,Lopez2006}). In what fallows, we consider the line centroid as a free parameter of the fit, 
the lower and upper bounds of which are set equal to 6.4 and 6.5 keV, respectively.

\begin{table*}
\centering
\begin{tabular}{cccc|cc}\hline\hline
 & & \multicolumn{2}{c}{Blue jet} & \multicolumn{2}{c}{Red jet}  \\
Spectroscopic &  $E_0$,  & $E_{zb}$, & Flux (Eq. Width), & $E_{zr}$, & Flux (Eq. Width), \\
symbol & keV & keV & $10^{-4}$ \ph\ (eV) & keV & $10^{-4}$ \ph\ (eV)\\
\hline
\rule{0pt}{3ex}
\feK  & 6.70 & 6.92 & $7.17_{7.08}^{7.26}$ ($379_{373}^{385}$) & 6.13 & $3.76_{3.70}^{3.83}$ ($200_{196}^{203}$) \\
\rule{0pt}{3ex}
\feKb  & 7.88 & 8.14 & $0.6_{0.4}^{0.7}$ ($40_{30}^{50}$) & 7.20 & ${\scriptstyle = F_{b}({\rm Fe\, XXV\ K}_{\beta}) \frac{F_r({\rm Fe\, XXV\ K}_{\alpha})}  {F_{b} ( {\rm Fe\, XXV\ K}_{\alpha} )}}$\\
\rule{0pt}{3ex}
\feL  & 6.97 & 7.20 & $2.68_{2.61}^{2.76}$ ($119_{116}^{123}$) & 6.37 & ${\scriptstyle =F({\rm Fe\, I\ K}_\alpha)}$ \\
\rule{0pt}{3ex}
\niK & 7.80 & 8.05 & $2.07_{2.02}^{2.13}$ ($181_{174}^{185}$) & 7.13 & ${\scriptstyle = F_b({\rm Ni\, XXVII\ K}_{\alpha}) \frac{F_r({\rm Fe\, XXV\ K}_{\alpha})}{F_b({\rm Fe\, XXV\ K}_{\alpha})}}$ \\ 
\rule{0pt}{3ex}
\niL & 8.10 & 8.36 & $0.87_{0.81}^{0.92}$ ($71_{67}^{75}$) & 7.41 & $=0$ \\
\hline\hline
\multicolumn{6}{c}{Fluorescent lines} \\
Spectroscopic & $E_0$, & \multicolumn{4}{c}{Flux (Eq. Width),}  \\
symbol & keV & \multicolumn{4}{c}{$10^{-4}$ \ph\ (eV)} \\
\hline
\rule{0pt}{3ex}
\fe  &  $6.450_{6.441}^{6.456}$  & \multicolumn{4}{c}{$1.11_{1.07}^{1.15}$ ($53_{51}^{55}$)} \\
\rule{0pt}{3ex}
\nikel  & $7.500_{7.500}^{7.505}$  & \multicolumn{4}{c}{ $0.26_{0.22}^{0.30}$ ($16_{14}^{18}$) }  \\
\hline
Fe\,(\textsc{\lowercase{XXII--XXIII}}\,K$_{\alpha}$?) & $6.63_{6.62}^{6.64}$  & \multicolumn{4}{c}{ $0.98_{0.92}^{1.04} (39_{36}^{41})$ } \\
\end{tabular} 
\caption{\small
The set of simulated lines within the phenomenological method of data description. The centroid energies $E_0$ are the weighted mean energies of the corresponding triplets and doublets in the rest frame of the jet. The indicated best fit parameters are given for fitting
of a sum of lines and bremsstrahlung continuum with a temperature $T = 22.3_{21.5}^{23.2}$ keV and a flux of $F_{6-9} =  4.22_{4.20}^{4.24} \times 10^{-11} $ \ergscm\ to the 120 ks spectrum in the 4.3--12 keV energy range. The data was obtained with the {\textit XMM-Newton} EPIC-pn camera. The ranges of the parameter values correspond to 90\% confidence level (see details in Appendix~\ref{sec:append}). 
The positions of the shifted lines $E_{zb}$ and $E_{zr}$ are calculated from the best fit parameters of Doppler shifts: $z_b = -3.15^{-3.13}_{-3.18}\times 10^{-2}$ and $z_r = 9.38^{9.43}_{9.33}\times 10^{-2}$, for the blue and red jets respectively. The fluxes in the lines are indicated in the emitter's rest frame without taking into account the flux in the continuum. The fluxes of the red jet lines \feKb\ and \niK\ are determined through the fluxes of the corresponding lines of the blue jet. The flux of the red jet line \feL\ is assumed to be equal to the flux of the \fe\ line. The \niL\ line of the red jet is not included. The width of the jets lines is found to be equal to $\Sigma_{E0, jet} =76_{74}^{78}$ eV, the width of the fluorescent lines is fixed: $\Sigma_{{\rm Ni\,I}} = \Sigma_{{\rm Fe\,I}}  = 5$ eV (see the text for more details). The notation ``Fe\,\textsc{\lowercase{XXII--XXIII}}\,K$_{\alpha}$'' shows an unidentified line with a width of $\Sigma < 5$ eV, which must be added to the model to avoid significant residuals in the energy region of 6.6 keV (see Fig.~\ref{f:bremss_gauss}). The fit quality corresponds to $\chi^2/{\rm d.o.f} = 2174 / 1526 =  1.42$ (criterion $\chi^2$ to the number of degrees of freedom).
}
\label{t:lines}
\end{table*}

The ``motion'' of the jet emission lines leads to their mutual overlapping (blending) in the spectrum.
As can be seen from Fig.~\ref{f:eefluor} in Section~\ref{sec:data}, for a characteristic line width corresponding to 
the EPIC-pn spectral response function, the line blending appears throughout most of a precessional cycle of SS~433 (see also $E_{zb}$ and $E_{zr}$ in Table~\ref{t:lines}). Within the framework of the phenomenological model under consideration, this leads to a degeneracy of weak line parameters, which are blended with brighter lines in the spectrum. 
In order to place constraints on the flux of such lines, we introduce a number of simplifying assumptions:
\begin{enumerate}
\item  the flux of the red jet \niK\ line is fixed equal to the flux of the corresponding blue jet line multiplied by 
the flux ratio of the red and blue  \feK\  lines:
$${\small F_r({\rm Ni\,XXVII\,K}_{\alpha}) = F_b({\rm Ni\,XXVII\,K}_{\alpha}) \frac{F_r({\rm Fe\,XXV\,K}_{\alpha})}{F_b({\rm Fe\,XXV\,K}_{\alpha})}};$$
\item the flux of the red jet \feKb\ line is defined in a similar way:
$${\small F_r({\rm Fe\,XXV\,K}_{\beta}) = F_b({\rm Fe\,XXV\,K}_{\beta}) \frac{F_r({\rm Fe\,XXV\,K}_{\alpha})}{F_b({\rm Fe\,XXV\,K}_{\alpha})}};$$
\item 
the red jet \feL\ line and the \fe\ fluorescent line are blended. In this case, it is difficult to constrain the \feL\ flux by the flux of the corresponding blue jet line, since emissivity of the Ly$_{\alpha}$ doublet, corresponding to the transitions in hydrogen-like iron, has a peak at higher temperatures than in the case of helium-like iron K$_{\alpha}$ triplet. Testing different variants of the fit, we came to the conclusion about the expediency of fixing fluxes of the blended lines equal to each other:  
$${\small F_r({\rm Fe\,XXVI\,Ly}_{\alpha}) = F({\rm Fe\,I\,K}_{\alpha})}.$$ The \fe\ line flux obtained in this way is well consistent with the results of fitting of the \texttt{lbjet} model (see the next subsection), and also with the findings of \cite{Lopez2006} based on \textit{Chandra} data;
\item 
the red jet \niL\ line fall in the vicinity of the investigated \nikel\ line. In view of the above uncertainty with respect to $F_r({\rm Fe\,XXVI\,Ly}_{\alpha})/F_b({\rm Fe\,XXVI\,Ly}_{\alpha})$ , it is also hard to constrain the flux $F_r({\rm Ni\,XXVIII\,Ly}_{\alpha})$ using the flux of the corresponding blue jet line. Since we aim to find weak spectral features at 7.5 keV, 
this line is excluded from the fitting model.
\end{enumerate}

Next, we add a narrow Gaussian corresponding to the \nikel\ line with a centroid at $E_{{\rm Ni\, I}} = 7.5$ keV and width of $\Sigma_{{\rm Ni\,I}} = \Sigma_{{\rm Fe\,I}}  = 5$ eV. Same as for the \fe line, we consider the line centroid as a free parameter with a range from 7.5 to 7.6 keV.

The fit of the described model to the observed spectrum leaves significant residuals in the 6.6--6.7 keV energy range.
Probably, it can be related to a spectral line (or a complex of lines) unaccounted for in the model.
For the obtained best-fit Doppler shifts, no emission lines of the jets fall into this range. 
However, if the different parts of the accretion disc wind have different ionization degrees, then the lines 
associated with iron fluorescence from the highly ionized part of the wind can fall in this range \citep[see, e.g.,][]{Kallman2004}.
To improve the fit, we then add to the model a narrow Gaussian line with a width of $\Sigma < 5$ eV and a centroid at 6.6 keV. 
The obtained line best-fit parameters correspond to the centroid energy $E_0 = 6.63_{6.62}^{6.64}$ keV and the flux of $0.98_{0.92}^{1.04} \times 10^{-4}$  \ph. The line flux is well constrained from spectral fitting. 
The centroid energy is consistent with the ionization state of iron up to \textsc{\lowercase{XXII-XXIII}}. 
In what follows, we will not dwell on the description of this line, since the EPIC-pn energy resolution does not allow investigating the line profile and drawing any conclusions regarding its origin.

The results of the fitting are summarized in Table~\ref{t:lines}, the best-fitting model is shown in Fig.~\ref{f:bremss_gauss}. The total number of free parameters of the model is 15: the fluxes of the jet lines (6), their  width  $\Sigma_{E0, jet}$ (assumed equal for all jet lines), the line Doppler-shifting for the blue and red jets ($z_b$ and $z_r$), the centroids and fluxes of the fluorescent and
  unidentified lines (6).

\subsection{X-ray emission from baryonic jets}
We now proceed from the phenomenological description of the strongest observed lines to the 
global fitting of the entire line complex arising from the thermal emission of the jets.
To this end, we use the \texttt{lbjet} model described in Section~\ref{sec:models}. 
As noted above, self-consistent analysis of spectral lines and continuum is beyond the scope of this study. We  therefore
use the same continuum model as before with parameters inferred from the 4.3--5.8 keV and 10--12 keV energy bands, which do not contain bright lines (see Section~\ref{sec:bremss}).

The \texttt{lbjet} model inherits the parameter space of the \texttt{bjet} model \citep[for details, see Section~\ref{sec:models} and][]{KMS2016}. However in the context of the current study, only
the gas temperature at the jet base $T_0$ and relative nickel abundance $Z_{{\rm Ni}}/Z$ become essential parameters of the model.
The abundance of other heavy elements, as in the \texttt{bjet} model, is given by the parameter $Z$ relative to the set of solar abundances of \cite{Anders1989}.   Assuming that cooling of the SS~433 jets near the base is dominated by adiabatic expansion  ($\alpha<1$, see \citealt{KMS2016}),  
the parameters of kinetic luminosity $L_k$ and optical depth for electron scattering at the jet base  $\tau_{e0}$ 
make sense only as the normalization-defining combination $L_k\tau_{e0}$ of the spectral lines.
Therefore the parameter $Z$ is degenerate and is not used during the fitting procedure, so that we fix $Z=1$. 
Moreover, we assume equal temperatures of the two jets in order to avoid degeneracy between the flux of the \fe\ line (due to blending with the \feL\ line) and the temperature of the red jet.  
We have tested this assumption by decoupling the temperatures of the jets in the fitting model.  
We found that the difference between the best-fit values of the jets temperatures  does not exceed 2 keV, i.e. it is relatively small.
The assumption needs to be invoked when introducing a continuum model with a neutral iron absorption edge (see Section~\ref{sec:edge}), since
it appears that the photoabsorbed flux can be compensated by an increase in the temperature of the red jet.
Consequently, the flux of the red jet \niL\ line will increase with decreasing in flux of the \fe\ line.
As a result, the temperature of the red jet can become significantly higher than one of the blue jet. 
On the basis of general ideas about the geometry of the system, it is obvious that such a range of parameter values is physically unjustified.

\begin{table}
\small
\centering
\begin{tabular}{ccc}\hline\hline
\texttt{lbjet}  & Blue jet & Red jet  \\
parameter & & \\
\hline
\rule{0pt}{3ex}
$T_0$, keV &  $12.3_{11.9}^{12.6}$  & $=T_{0,b}$ \\
\rule{0pt}{3ex}
$Z_{{\rm Ni}}/Z$ & $8.9_{8.7}^{9.3}$  & $=(Z_{{\rm Ni}}/Z)_b$ \\ 
\rule{0pt}{3ex}
$n = \tau_{e0} \times L_k^a$ & $3.72_{3.66}^{3.76}$  & $1.80_{1.77}^{1.83}$  \\
\rule{0pt}{3ex}
$z \times 10^{-2}$ &  $-3.51_{-3.53}^{-3.49}$  & $8.98_{8.93}^{9.03}$ \\
\rule{0pt}{3ex}
$\Sigma_{jet}$, eV &  $63_{61}^{65}$  & $=\Sigma_{jet,b}$ \\
\hline\hline
\multicolumn{3}{c}{Fluorescent lines} \\
Spectroscopic & $E_0$, & Flux (Eq. width),  \\
symbol  & keV & $10^{-4}$ \ph\ (eV) \\
\hline
\rule{0pt}{3ex}
\fe & $6.449_{6.440}^{6.456}$  & $0.96_{0.94}^{1.00}$ ($46_{44}^{48}$) \\
\rule{0pt}{3ex}
\nikel & $7.5_{7.5}^{7.5}$  &  $0.00_{0.00}^{0.01}$ ($0.0_{0.0}^{0.9}$)   \\
\rule{0pt}{3ex}
Fe & $6.63_{6.62}^{6.64}$  & $0.97_{0.93}^{1.04} (39_{36}^{41})$  \\
(\textsc{\lowercase{XXII-XXIII}}\,K$_{\alpha}$ ?) & & \\
\hline
\multicolumn{3}{l}{$\ ^{a}\ \times 10^{38}$ erg/s }
\end{tabular} 
\caption{\small
Results of fitting of the \texttt{lbjet} model to the \textit{XMM-Newton} EPIC-pn spectrum  of SS~433 with 120 ks exposure.
$T_0=T_{0,b}=T_{0,r}$ is the jets base temperature, $Z_{{\rm Ni}}/Z$ is the relative abundance of nickel in the jets (in solar units), $n$ is normalizing parameter of the jet line intensities. The line width  $\Sigma_{jet}$ is set equal for the blue and red jets. The notation ``Fe\,\textsc{\lowercase{XXII--XXIII}}\,K$_{\alpha}$'' shows an unidentified line (see Section~\ref{sec:gauss}). The quality of the fit corresponds to $\chi^2/{\rm d.o.f} = 2231/1528 = 1.46$.}
\label{t:lbjet}
\end{table}

The line broadening due to the transverse velocity of the jets is set by the convolution of  \texttt{lbjet} model and  Gaussian function with a width $\Sigma_{E0} = \Sigma_{jet} \times (E_0/$6 keV).  The line Doppler shifting, as before, is determined by the parameters $z_b$ and $z_r$. The \fe\ and \nikel\ fluorescent lines are described by narrow Gaussians with widths $\Sigma_{{\rm Ni\,I}} = \Sigma_{{\rm Fe\,I}}  = 5$ eV. In addition, similar to the previous model, we add a narrow Gaussian with a width $\Sigma < 5$ eV and a centroid at 6.6 keV. The total number of free parameters of the model is 13.

The obtained  best-fit parameters are presented in Table~\ref{t:lbjet}. 
The {\texttt lbjet} model has an important advantage over the phenomenological method in that 
the relative fluxes of weak lines can be constrained by the jet temperature parameter, so  that
the simplifying assumptions described in Section~\ref{sec:gauss} are no longer needed. 
As can be seen from the obtained \nikel\ line flux, the contribution of the red jet lines, which are unaccounted for in the phenomenological model, turns out to be important when fitting the spectrum near the 7.5 keV energy.
The major contributors to this energy region are  \niL\  and the weaker lines of the K-series ($\beta$,$\gamma$,$\delta$...) of helium-like iron of the red jet\footnote{Satellites of bright lines can make a significant contribution as well \citep[see details in][]{KS2012}}.

The best-fit value for the width of the jet lines is found to be $\Sigma_{jet}=63_{61}^{65}$ eV, which is more than twice the value inferred from {\it Chandra} data in the 1--3 keV soft X-ray band ($\Sigma_{jet} \sim 30$ eV, which corresponds to the opening angle of $\sim 1.5^{\circ}$ \citealt{Marshall2002,KMS2016}). 
On the one hand, the jet lines in the 6--9 keV band were actually observed to be more broadened than those in the softer energy band.
In particular, for \feK\ triplet {\it Chandra}/HETGS measurements result in $\Sigma_{E0} \sim 50$ eV  \citep{Namiki2003}.
On the other hand, the \textit{XMM-Newton} EPIC-pn spectrometer has a relatively low spectral resolution $\Delta E_{FWHM} \gtrsim 2.36 \Sigma_{jet}$ resulting in overlapping of close lines.
In addition, the spectral variability of the source on the time scale shorter than the total exposure (due to, e.g., the nodding or jitter, see next subsection) can also contribute to the broadening of the observed lines.

\subsection{Spectral evolution}
\label{sec:10ks}
As can be seen from the light curves (see Fig.~\ref{f:light_curve}), the flux in the hard band of the standard X-ray energy range exhibits a noticeable variability on the time scale of the total time of observation. In addition, the spectral evolution of the source is also significant. To examine the spectral variability, we break the one-day data with the 120 ks total exposure into 12 parts of 10 ks duration and perform a spectral fitting to each spectrum.

\begin{figure}
\centering
\includegraphics[width=1\columnwidth]{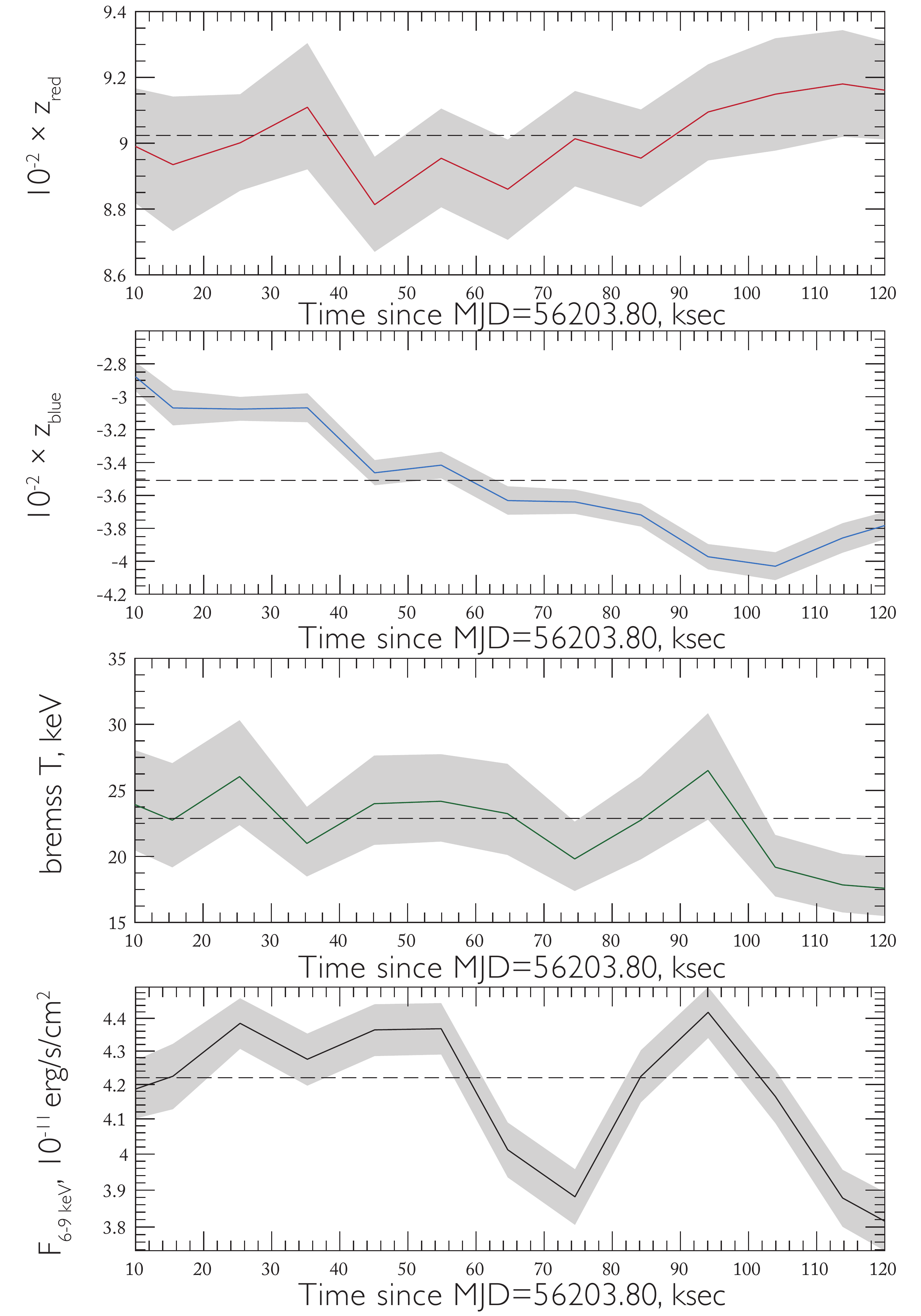}
\vspace{-10pt}
\caption{\small
The spectral variability of SS~433 determined from the \textit{XMM-Newton} data with the total exposure of 120 ks. 
The one-day data was divided into 12 time-sliced spectra with a regular interval of 10 ks.
The continuum is described by fitting of thermal bremsstrahlung to each spectrum in the 4.3--5.8 keV and 10--12 keV energy ranges, which do not contain bright lines. The best-fit parameters of the continuum model are 
shown in the two bottom panels: bremsstrahlung temperature and flux in the 6--9 keV range.
The two top panels show the best-fit values of Doppler shifts for fitting of the \texttt{lbjet} model to each spectrum in the 4.3--12 keV range.
The gray bands show  the 90\% confidence range. The dashed horizontal lines show the average 
values determined from the time-averaged spectrum.}
\label{f:dz}
\end{figure}

The Doppler shifts of the lines of red and blue jets obtained from the fitting of the \texttt{lbjet} model  to individual 10 ks spectra are shown in the two top panels of Fig.~\ref{f:dz}. The bottom panels show the corresponding parameters of the continuum model. The averaged over 12 spectra quality of the continuum fit corresponds to $\chi^2/{\rm d.o.f} = 700 / 696 = 1.006$ (we remind that unbinned weighted spectra are used in this work).

The average rate of change in the position of the blue jet lines during first 100 ks from start of exposure corresponds to  $\Delta z_b/ \Delta t = 0.99_{0.85}^{1.11} \times 10^{-2}$ d$^{-1}$.
The most rapid variability of the blue jet Doppler shift on the time scale of 10 ks is detected at the level of $\Delta z_b \approx 0.4 \cdot 10^{-2}$, which corresponds to ${\rm max}(dz_b/dt) = 3.45_{2.01}^{4.89} \times 10^{-2}$ d$^{-1}$. 
At the same time, the maximum variability due to the sinusoidal nodding motion of the jets with a 6.3 d period is expected at the level of $dz/dt \approx 1.21  \times 10^{-2}$ d$^{-1}$ \citep{Fabrika2004}. 
Thus, the average change in the position of the lines on the time scale of the total exposure is consistent with the nodding-precession variability of the system, while the rapid change 
 in a time of 10 ks is larger and can be related to the jitter motions of the jets \citep[see][]{Iijima1993,Kubota2010}. If we now assume that the line broadening  due to the jets transverse velocity component is $\Sigma_\Theta \approx 30 \times (E_0$/6 keV) eV, we can estimate the broadening of the lines due to the jitter:
\begin{equation}
\Sigma \approx \sqrt {\Sigma_\Theta^2 + (\delta E_0 / 2 \sqrt{3})^2},
\end{equation}
where $\delta E_0  =  \frac{E_0} {(1 + z)^2} \Delta z $ --- the change in the energy of the line centroid in 10 ks time interval. For the \feK\ line (6.4 keV) of the blue jet and ${\rm max}(\Delta z_b) = 0.39 \times 10^{-2}$, we obtain $\delta E_{0,Fe\,XXV\,K\alpha} = 26.9$ eV. In this case, the line broadening due to the jitter should not be more than $5$ \% of the width of $\Sigma_\Theta$. At the same time, the jet line width averaged over 12 parts corresponds to $\Sigma_{jet} = 53_{47}^{59}$ eV, which is consistent with the findings of \cite{Namiki2003} within the uncertainties.

\begin{figure}
\centering
\includegraphics[width=1\columnwidth]{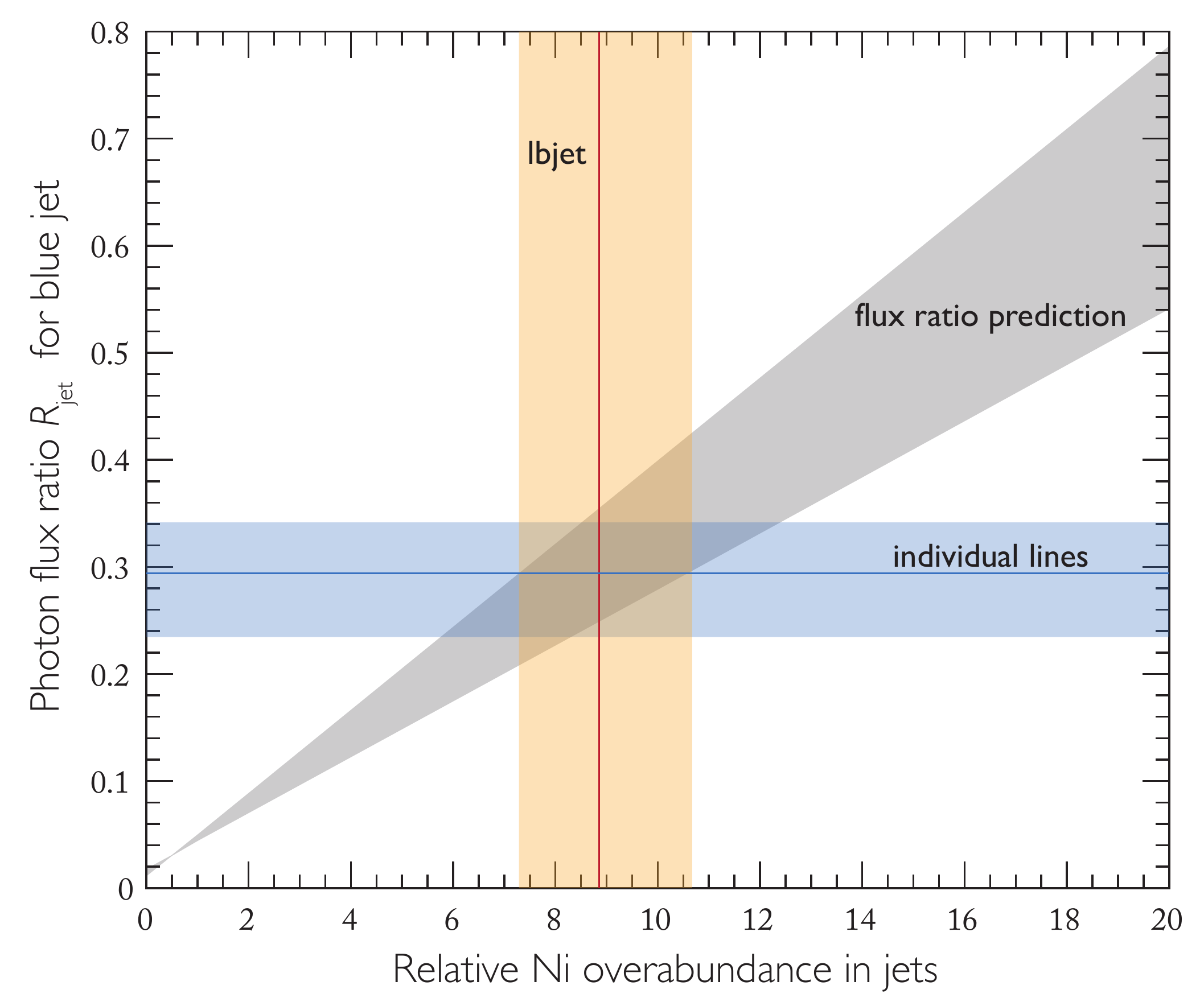}
\vspace{-10pt}
\caption{\small
Relative nickel abundance in the jets of SS~433 determined from \textit{XMM-Newton} data. 
The total 120 ks exposure was divided into 12 equal parts, the shown results correspond 
to the average values of the best-fit parameters for each 10 ks spectrum.
The red line shows the relative nickel abundance $Z_{{\rm Ni}} / Z$, determined 
from fitting of \texttt{lbjet} model. 
The blue line shows the photon flux ratio of K$\alpha$-triplets of helium-like nickel and iron for the blue jet, $R_{{\rm jet}} = F({\rm Ni\,XXVII\,K}\alpha) / F({\rm Fe\,XXVII\,K}\alpha)$, obtained by fitting of the phenomenological model (marked as ``individual lines'').
The orange and blue areas show the 90\% confidence ranges for  $Z_{{\rm Ni}} / Z$  and  $R_{{\rm jet}}$, respectively. Prediction of the photon flux ratio $R_{{\rm jet}}$ in the \texttt{lbjet} model
as a function of the nickel abundance is depicted by the gray area with the scatter coming from dependence on the other model parameters (mainly temperature).
}
\label{f:jetNi}
\end{figure}

To exclude possible influence of the source spectral variability on the derived parameters, we present the results of fitting of the phenomenological and  \texttt{lbjet} models to each 10 ks spectrum
in  Table~\ref{t:results_noedge} for the flux of fluorescent lines and the relative nickel abundance $Z_{{\rm Ni}} / Z$.  Within the framework of the phenomenological model, the nickel abundance $Z_{{\rm Ni}} / Z$  is found from the flux ratio of the brightest lines of nickel and iron in the spectrum: $R_{{\rm jet}} = F({\rm Ni\,XXVII\,K}\alpha) / F({\rm Fe\,XXVII\,K}\alpha)$. 
In Fig.~\ref{f:jetNi} this value is shown by the blue horizontal line (the blue area corresponds to the 90\% confidence range). We also calculated the $R_{{\rm jet}}$ value using the \texttt{lbjet} model 
across the range of parameters $\tau_{e0}$ from $5 \cdot 10^{-5}$ to 0.5 and $T_0$ from 7 to 40 keV, which corresponds to the range of $\alpha$ from $10^{-4}$ to 10. 
The obtained prediction is shown in Fig.~\ref{f:jetNi} by the gray area (see also Fig.~6 in \citealt{KMS2016}). Value $Z_{{\rm Ni}} / Z$  in Table~\ref{t:results_noedge} for the phenomenological model corresponds to the intersection of the blue and gray areas. The best-fit value of $Z_{{\rm Ni}} / Z$ of the \texttt{lbjet} model is shown with red vertical line (orange area --- 90\% confidence range). The relative abundance of nickel obtained for two models on the whole is in good agreement (by the average value), which serves as an additional argument in favor of the adequacy of the approximation of the data by the \texttt{lbjet} model, since in this case the parameter of the nickel abundance depends not only on the flux ratio $R_{{\rm jet}}$, but also on the jet temperature, determined by the fitting to all lines in the spectrum in the energy range under consideration.

\begin{table}
\small
\centering
\begin{tabular}{ccc}\hline\hline
 &   Phenomenological  & \texttt{lbjet}\\
& model &  \\
\hline
\rule{0pt}{3ex}
$F({\rm Fe\,I\ K}_\alpha)$,  &   $1.11_{0.96}^{1.13}$ &   $0.99_{0.84}^{1.12}$\\
$10^{-4}$ \ph & & \\ 
\rule{0pt}{3ex}
$ F({\rm Ni\,I\ K}_\alpha)$,  & $0.21_{0.07}^{0.35}$   & $0.03_{0.00}^{0.09}$  \\
$10^{-4}$ \ph & & \\
\rule{0pt}{3ex}
$R_{{\rm fluor}}$,  &  $0.17_{0.05}^{0.29} $  &  $0.03_{0.00}^{0.10} $\\
\rule{0pt}{3ex}
$Z_{{\rm Ni}}/Z$ & $8.6_{5.8}^{12.3}$ & $8.6_{7.3}^{10.3}$ \\ 
\rule{0pt}{3ex}
$Z_{{\rm Ni, wind}}/Z_{{\rm Ni, jet}}$ & $0.44_{0.16}^{0.64}$ & $0.08_{0.00}^{0.24}$   \\ 
\rule{0pt}{3ex}
$\chi^2/{\rm d.o.f}$ &   $1544 / 1526$ &  $1566 / 1528$ \\
\hline 
\end{tabular} 
\caption{
Fluxes of the \nikel\ and \fe\ fluorescent lines, their ratio $R_{{\rm fluor}}$, the relative nickel abundance in the jets  $Z_{{\rm Ni}}/Z$, and the ratio of the abundances of nickel in the wind and in the jets of SS~433 $Z_{{\rm Ni, wind}}/Z_{{\rm Ni, jet}}$, obtained by a sum 
of bremsstrahlung continuum (without an absorption edge from neutral iron described in Section~\ref{sec:edge}) and phenomenological or \texttt{lbjet} model. The listed values correspond to average values of the best-fit parameters for 10 ks spectra in the 4.3--12 keV energy range.}
\label{t:results_noedge}
\end{table}

As can be seen from Table~\ref{t:results_noedge}, in addition to the overestimation of $R_{{\rm fluor}}$ value described in the previous subsection (due to  unaccounted red \niL line), the phenomenological model gives a larger scatter in the $Z_{{\rm Ni}}/Z$ value. This result is a consequence of the large range of possible values of the parameter $Z_{{\rm Ni}}/Z$ for a fixed flux ratio $R_{{\rm jet}}$  (gray area in Fig.~\ref{f:jetNi}).

\subsection{Neutral iron absorption edge}
\label{sec:edge}
At the final step, we investigate how measurements of
the flux ratio $R_{{\rm fluor}} = F({\rm Ni\,I\ K}_{\alpha})/F({\rm Fe\,I\ K}_{\alpha})$
can be affected by  the presence of
 the absorption edge of neutral iron at 7.1 keV. For that purpose, we describe the continuum model with the absorption edge as follows:
\begin{equation}
I(E) = 
\begin{cases}
I_{ff}(E) & E \leqslant E_{{\rm edge}} \\
I_{{\rm edge}}(E) I_{ff}(E)&  E>E_{{\rm edge}}
\end{cases}
,
\label{eq:edge}
\end{equation}
where $I_{ff}(E)$  is the bremsstrahlung spectrum, $E_{{\rm edge}} = 7.1$ keV is threshold energy
 and $I_{{\rm edge}}(E) = \widetilde{N} \exp [ -\tau_{{\rm edge}} (\frac{E}{E_{{\rm edge}}})^{-3}] $  is the absorption edge model, parametrized by two quantities: $\tau_{{\rm edge}}$ is the optical depth for photoabsorption at the energy $E_{{\rm edge}}$ and $\widetilde{N}$ is the normalization parameter, which determines the fraction
of photons being absorbed by neutral iron along the line of sight.  
 As can be seen from Section~\ref{sec:models}, the shape and depth of the absorption edge depend on the geometry and the optical depth of the reflecting/scattering medium. To cover a wide range of possible scenarios, we define the parameter $\widetilde{N} $ through the flux of \fe\ fluorescent line, so that the following equality is fulfilled:
\begin{equation}
\small
N_{{\rm edge}}\, F({\rm Fe\,I\ K}_\alpha) = \int\limits_{E_{{\rm edge}}}^{\infty} [ I_{ff}(E) -  I_{{\rm edge}}(E)  I_{ff}(E) ] dE,
\label{eq:Nedge}
\end{equation}
where $N_{{\rm edge}}$ is model parameter.

\begin{figure}
\centering
\includegraphics[width=1.05\columnwidth]{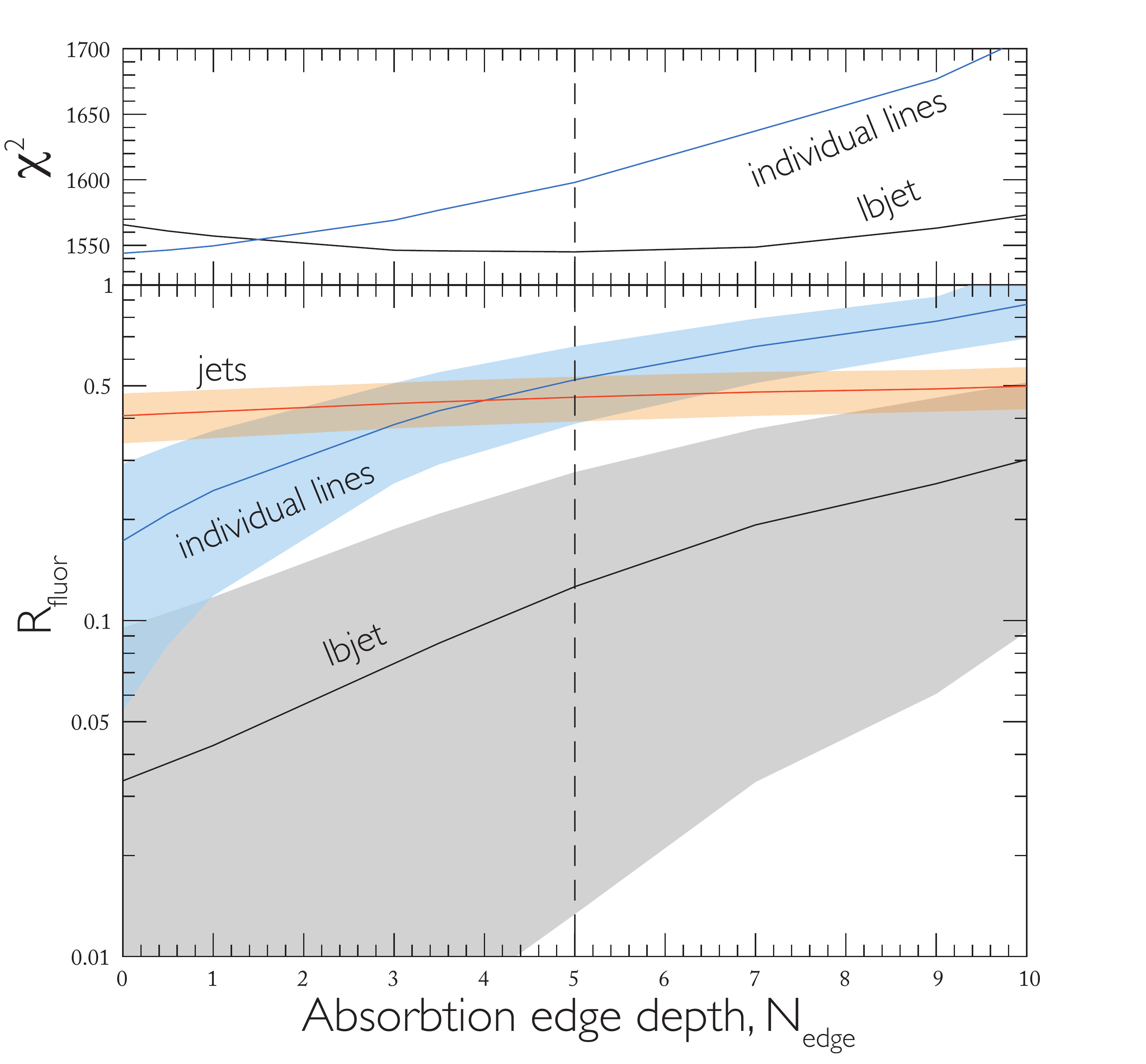}
\vspace{-10pt}
\caption{\small
Bottom panel. The photon flux ratio of \nikel\ and \fe\ fluorescent lines 
 as a function of the depth parameter $N_{{\rm edge}}$ of the neutral iron absorption edge model. The parameter $N_{{\rm edge}}$ is defined as the fraction of absorbed photons along the line of sight in units of the photon flux of \fe\ fluorescent line (equations~\ref{eq:edge} and~\ref{eq:Nedge}).  The $N_{{\rm edge}}$ parameter is fixed during the fit on a grid of values from 0 to 10.
 The flux ratio shown is the average value of the best-fit parameters 
of the phenomenological (blue line) and  \texttt{lbjet} (black line) models for 
 10 ks \textit{XMM-Newton} spectra. 
 The corresponding 90\% confidence ranges are depicted by the blue and gray areas, respectively. 
 The orange area corresponds to the predicted $R_{\rm fluor}$ ratio from the best-fit value of relative nickel abundance in the jets for the \texttt{lbjet} model: $R_{\rm fluor} = 0.045 \times Z_{{\rm Ni} } / Z$ (see Section~\ref{sec:models}). 
Top panel. The best-fit $\chi^2$-statistic achieved by the phenomenological (blue line, d.o.f$ = 1526$) and \texttt{lbjet} (black line,  d.o.f$ = 1528$) models for data in the 4.3--12 keV range
as a function of the $N_{{\rm edge}}$ parameter.
The vertical dashed line shows 
the value of $N_{{\rm edge}}$ parameter, for which 
the \texttt{lbjet} model achieves best $\chi^2$-statistic.}
\label{f:NiI_edge}
\end{figure}

We consider the optical depth at the absorption edge energy $\tau_{{\rm edge}}$ as a free parameter of the model. To estimate the upper limit on this parameter 
we adopt the value  $\sigma_{Fe\,I}=3.764 \times 10^{-20}$ cm$^{2}$ (\texttt{XCOM} database\footnote{https://www.nist.gov/pml/xcom-photon-cross-sections-database})
for the photoelectric absorption cross-section of neutral iron at the edge energy.
Then the ratio of the optical depth for the photoabsorption of neutral iron to the optical depth for electron scattering along the line of sight can be estimated as:
\begin{equation}
\left. \frac{\tau_{{\rm ph.abs}}}{\tau_e}\right|_{E=7.1{\rm keV}} \sim \frac{n_{Fe}}{n_e} \frac{\sigma_{FeI}}{\sigma_{e}}  \approx 2.2,
\end{equation}
for the solar element abundance \citep{Anders1989}.
Therefore, as a conservative estimate, we set the upper bound on this parameter to ${\rm max} (\tau_{{\rm edge}}) = 2$.

In view of the relatively poor spectral resolution of the EPIC-pn instrument, it is difficult to constrain the $N_{{\rm edge}}$ parameter by the data in hand, since   
the likelihood function varies only weakly along $N_{{\rm edge}}$ direction in parameter space of 
a model with the continuum given by equation~\ref{eq:edge}.
Degeneracies of the $N_{{\rm edge}}$ parameter leaves some parameter combinations poorly constrained as well, resulting in the overestimated errors (see Appendix~\ref{sec:append} and marginalized two-dimensional distributions in Fig.~\ref{f:margin}). 
In order to avoid large uncertainties on parameter values, we fix $N_{{\rm edge}}$ on a grid of values and perform the fitting procedure, described in the previous section, for each $N_{{\rm edge}}$ value (we show the best-fit model with  $N_{{\rm edge}}$  left  free in Appendix~\ref{sec:append}).

To get some insight regarding limitations on the $N_{{\rm edge}}$ parameter values,  
we carried out the following calculations.
Using the two models described in Section~\ref{sec:models}, \texttt{pexmon} and \texttt{cwind}, we 
calculated the emergent spectra for a broad range of model parameters.
The continuum of simulated spectra were then fitted by 
  a smooth function in the energy ranges which are not affected by the absorption edge and do not contain fluorescent lines. 
Next, the photon flux of the best-fit smooth function  above 7.1 keV was subtracted from 
 the flux of simulated spectra with the absorption edge. This makes it possible for 
  the value  corresponding to normalization $\widetilde{N}$ to be found. 
Similarly, we determined the predicted photon flux of the \fe\ line, which in the end allows us to calculate the relative normalization $N_{{\rm edge}}$ of the absorption edge, which we use to analyze the spectra.

\begin{table}
\small
\centering
\begin{tabular}{ccc}\hline\hline
 &   Phenomenological  & \texttt{lbjet}\\
& model &  \\
\hline
\rule{0pt}{3ex}
$F({\rm Fe\,I\ K}_\alpha)$,  &  $1.11_{0.96}^{1.13}$ &   $0.86_{0.72}^{1.11}$   \\
$10^{-4}$ \ph & & \\ 
\rule{0pt}{3ex}
$ F({\rm Ni\,I\ K}_\alpha)$,  &   $0.21_{0.07}^{0.35}$ & $0.11_{0.01}^{0.25}$ \\
$10^{-4}$ \ph & & \\
\rule{0pt}{3ex}
$R_{{\rm fluor}}$,  &  $0.17_{0.05}^{0.29} $ & $0.13_{0.01}^{0.21} $   \\
\rule{0pt}{3ex}
$Z_{{\rm Ni}}/Z$ &  $8.6_{5.8}^{12.3}$ & $10.1_{8.5}^{11.6} $  \\ 
\rule{0pt}{3ex}
$Z_{{\rm Ni, wind}}/Z_{{\rm Ni, jet}}$ & $0.44_{0.16}^{0.64}$   &  $0.28^{0.62}_{0.03} $  \\ 
\rule{0pt}{3ex}
$N_{{\rm edge}}$ &  0  & 5   \\ 
\rule{0pt}{3ex}
$\chi^2/{\rm d.o.f}$ &  $1544 / 1526$ &   $1545 / 1528$ \\
\hline 
\end{tabular} 
\caption{\small
Fluxes of the \nikel\ and \fe\ fluorescent lines, their ratio $R_{{\rm fluor}}$, the relative nickel abundance in the jets  $Z_{{\rm Ni}}/Z$, and the ratio of the abundances of nickel in the wind and in the jets of SS~433 $Z_{{\rm Ni, wind}}/Z_{{\rm Ni, jet}}$, obtained by a sum 
of the continuum model with the neutral iron absorption edge  and phenomenological or \texttt{lbjet} model. The listed values are the average values of the best-fit parameters for 10 ks \textit{XMM-Newton} spectra. The parameter $N_{{\rm edge}}$ specifies the depth of the absorption edge in units of photon flux of the \fe\ line.  Fitting procedure is performed for fixed $N_{{\rm edge}}$ on a grid of values from 0 to 10. The presented results correspond to the  $N_{{\rm edge}}$ values, for which the models give the best agreement with the observed spectra in the 4.3--12 keV energy range.}
\label{t:results}
\end{table}

In the scenario of the  fluorescent lines formation as the reflection of radiation from an optically thick slab, represented by the \texttt{pexmon} model, the $N_{{\rm edge}}$ parameter lies in the range 2--2.5 for the inclination angles between the normal to the slab and the line of sight,  $i$, from 5 to 90 degrees \citep[see also][]{Basko1978}. 
For the model of the radiation passing through an optically thin cloud (the \texttt{cwind} model), the $N_{{\rm edge}}$ parameter is determined almost exclusively by the geometry of the scattering region. Thus, for a source placed in the center of a spherically symmetric cloud ($\mu_d = 1$), the flux ratio of  \fe\ line to the K-edge absorbed flux is approximately equal to the fluorescent yield of iron ($\omega_K = 0.34$, \citealt{Bambynek1972}), which gives $N_{{\rm edge}} \approx 3$  \citep[see more details in][]{Sunyaev1998}. If, however, the column density of the scattering matter along the line of sight is greater than the average over the angle around the source, then the emergent spectrum will be characterized by weak fluorescent lines. Therefore, for the line of sight perpendicular to the axis of the disc ($\mu =0$), the $N_{{\rm edge}}$ parameter becomes larger with increasing the wind funnel opening angle $\Theta_d = \arccos\mu_d$, reaching $\sim 10$ for $\Theta_d \approx 70^\circ$. Note that for the simple geometry assumed in the \texttt{cwind} model, the wind funnel opening angles $\Theta_d \gtrsim 60^\circ$ are unlikely, since at the precessional phases  of the SS~433 system close to zero, the angle between the line of sight and the axis of the accretion disc is $57^\circ$ \citep[see][]{Fabrika2004}, so the observer would look inside the wind funnel. In this case, one could expect a much greater precessional variability of the system in the X-ray energy range. It is possible, however, that the physical picture of the formation of fluorescent lines can be much more complicated, for example, as in the case of a clumpiness of the wind or different ionization degrees of the scattering matter. In this work, the range of the $N_{{\rm edge}}$ parameter values is set from 0 to 10.

The dependence of the  $R_{{\rm fluor}}$ flux ratio on the depth of the absorption edge  $N_{{\rm edge}}$  is shown in Fig.~\ref{f:NiI_edge}. The best-fit parameters of the two models are given in Table~\ref{t:results}  for the $N_{{\rm edge}}$ value, at which the models give the best agreement with the observed spectrum.
As seen from Fig.~\ref{f:NiI_edge}, when adding the absorption edge to the continuum model,
fitting with the phenomenological model gets poorer.
However, the \texttt{lbjet} model run including the absorption edge results in a $\chi^2$-statistic improvement:
the weak jet lines together contribute significantly to the continuum in the 7--9 keV range, so that the best $\chi^2$-statistic is achieved for $N_{{\rm edge}}=5$. When increasing  $N_{{\rm edge}}$, in addition to the flux of the \nikel\ line, the best-fit nickel abundance in jets also increases. However, the variation of this parameter turns out to be noticeably smaller.

\section{Discussion} 
\label{sec:discussion}

The fitting results for the models containing continuum with and without the absorption edge (given by equations~\ref{eq:edge}~and~\ref{eq:Nedge}) are summarized  in Table~\ref{t:results_noedge}~and~\ref{t:results}, respectively.
The phenomenological model allows more freedom in data description, since the observed line fluxes are all free parameters.
However, for the spectral resolution of the data used in the work, such freedom is excessive, since line blending in the spectrum leads to necessity of the additional assumptions to constrain the model parameters. Such assumptions in themselves can be justified only on the basis of the supposed physical picture of the formation of spectral lines. This is particularly relevant for relatively weak lines in the spectrum. In this case, the principal contribution is played by the contribution of the line \niL, the flux of which is expected to be at the level of 20\% of the \feL\ flux for the relative overabundance of nickel $\sim 10$ (based on the jet emission model), that is $\sim 2 \times 10^{-5}$ \ph\  for the red jet. Such a flux is very small in comparison with the flux of the iron lines of the blue jet, but comparable to or greater than the expected flux of the \nikel\ fluorescent line. For the precession phase of the system under consideration, the centroid of the \niL\ line of the red jet hits 7.41 keV, that is, in the immediate vicinity of the line of interest. For this reason, it can be concluded that the difference in the value of $R_{{\rm fluor}}$ obtained by two methods is mainly due to the contribution of the \niL\ line of the red jet, and therefore it is the parameters obtained for the \texttt{lbjet} model that should be considered as the actual result of this work.

Taking the obtained results at their face value, one can come to a conclusion that a significant amount of nickel observed in the jets should be newly synthesized somewhere in vicinity of the jets' launching pointing. On one hand, this might be considered as an indication for a neutron star as SS~433's compact object. Indeed, the surface of a supercritically accreting neutron stars appears as the most natural site for  
\textit{rp}-capture process of hydrogen burning that can lead to production of significant amount of nickel \citep[see, e.g.,][]{Schatz2001}. On the other hand, this almost necessarily implies, that we actually see  radioactive isotope nickel-56, that should half-decay into radioactive cobalt-56 in 6.1 days and than into a stable iron-56 in 77.1 days \citep[see, e.g.,][]{Nadyozhin1994}.
Note that the lines associated with the decay of $^{56}$Co $\rightarrow$ $^{56}$Fe were actually observed in supernovae of Type Ia 
with the INTEGRAL gamma-ray observatory \citep{Churazov2014}.

Because these decays are accompanied by emission of gamma-photons with $ \sim 1$ MeV energies, one might expect stable flux of gamma-ray lines that would be relatively narrow in case of $ ^{56}$Ni decay and significantly broadened for $ ^{56}$Co  as a result of averaging over period just twice smaller than precession period of the jets (162 days). The expected photon flux in these lines can be readily estimated assuming kinetic power of each jet at the level of $L_k \sim 3\times 10^{39}$ \ergs\ \citep{KMS2016}. Indeed, such kinetic power implies mass flow in the jet at the level of $\dot{M}_j=1.6\times10^{-6}~M_{\odot}$\,yr$^{-1}$, what corresponds to $ 10^{39} $  nuclei of $^{56}$Ni per second being permanently ejected and \textit{decaying} in the system. For the distance to SS~433 $ d\sim 5 $ kpc, this implies line photon flux of at most $ \sim 3.5\times10^{-7}$ \ph\ both for nickel and cobalt decay lines. Combining fluxes from the two jets one gets photon flux at the level of $\lesssim 10^{-6}$ \ph.

Such a line photon flux is at least two orders of magnitude less than the levels accessible with the \textit{INTEGRAL/SPI} spectrometer \citep{Winkler2003},  the most sensitive instrument currently available in the MeV-range.
Unfortunately, gamma-spectrometry in this spectral range is severely hindered by very high level of particle background, so the detection of the predicted signal will stay challenging even for the next generation of gamma-ray spectrometers 
such as a planned space mission \textit{e-ASTROGAM} with calorimeters \citep{Angelis2017}.

\section{Conclusions} 
\label{sec:conclusions}
In this paper, we have set an upper limit on nickel overabundance in the supercritical accretion disk wind of SS~433 from  
data of \textit{XMM-Newton} observation on October 3--5, 2012. 
The observation has been selected as the best suitable for searching for the  fluorescent line \nikel\ at 7.5 keV,
taking into account the sensitivity, brightness and precession phase of SS~433 (see Section~\ref{sec:data}).

We have considered two models of the formation of fluorescent lines in the X-ray spectrum of SS~433: 
 as a result of the reflection of hard X-rays from an optically thick neutral matter (Section~\ref{sec:reflection})
 and  as the scattering of the radiation by an optically thin gas followed by emission of fluorescent lines (Section~\ref{sec:scattering}).
 The first model can be related to 
the scenario, in which the illuminating radiation comes from 
the putative central X-ray source, to be reflected on the walls of the SS~433 accretion disk ``funnel'' \citep{Medvedev2010}.  
 As an approximation to the spectrum of reflected component arising in such a model, we use the \texttt{pexmon} model  \citep{Nandra2007}, that self-consistently reproduce fluorescent lines along with the reflected continuum.
In the second model, the additional hard component in the SS~433 spectrum is attributed to 
 the emission of the hottest parts of the jets, 
that are seen through  the optically thin for electron scattering, but optically thick for photoabsorption at energies below 3 keV,  wind of the supercritical accretion disk.  In this case, no bright central source needs to be invoked since emission in the hard band (above 4 keV) is not dramatically attenuated, although this component doesn't make any contribution in the soft band due to steep increase of the photoabsorption cross-section. The fluorescent line of iron here arises along with the Thomson scattered continuum.
 As an approximation of the emergent spectrum in this case, we have performed the Monte Carlo calculations using the spectral model \texttt{cwind} \citep{Churazov2017b}, which accounts for elastic and inelastic scattering, photo absorption of X-ray photons 
and fluorescence from neutral atoms of the most abundant heavy elements. 
The main results of Section~\ref{sec:models} are presented in Fig.~\ref{f:niferat}, where we show the predicted flux ratio of the fluorescent lines of iron and nickel: $R_{{\rm fluor}} = F({\rm Ni\,I\ K}_{\alpha})/F({\rm Fe\,I\ K}_{\alpha})$. 
Interestingly, the $R_{{\rm fluor}}$ value is obtained to be in the range of 0.45--0.5 for the nickel overabundance $Z_{{\rm Ni}}/Z = 10$ (in solar units of \citealt{Anders1989}) and almost independent of the other parameters of the models. 
We use the \texttt{cwind} model to verify that the $R_{{\rm fluor}}$ ratio is linearly scaled with the 
relative nickel-to-iron  abundance in the reflecting/scattering medium.

In Section~\ref{sec:results}, we have put a constraint on the fluorescent line flux ratio  $R_{{\rm fluor}}$ from the observational data. The data analysis was performed by two methods: 
1) using the phenomenological model, consisting of individual bright lines described in a form of a Gaussian; 
2) using the \texttt{lbjet} model, which is the continuum-subtracted version of the \texttt{bjet} model for
 thermal X-ray emission from a baryonic jet \citep{KMS2016}.
Since there is still no self-consistent model describing the observed spectrum of SS~433 in the 6--9 keV energy band,
we fit the continuum as bremsstrahlung emission (independent of the line components) in the 4.3--5.8 keV and 10--12 keV ranges,
that do not contain bright lines. The both methods give a similar result for the description of the  strongest spectral lines.
However, due to line blending, it is not possible to constrain fluxes of weak lines
 within the framework of the phenomenological model. Such weak lines of the red jet fall
in the vicinity of the \nikel\ line, making the flux of the latter poorly constrained.
Therefore, the method of global line fitting with the \texttt{lbjet} model is chosen as the main one.

During the \textit{XMM-Newton} observation a significant X-ray variability of SS~433 was detected. 
To exclude possible influence of such variability on the derived line fluxes,
we have broken the 120 ks data into 12 parts of 10 ks and have performed a spectral fitting to each part.
The fluorescent line fluxes averaged over the individual 10 ks spectra are obtained to be $F({\rm Ni\,I}) = 0.03_{0.00}^{0.09} \times 10^{-5}$ \ph\ and $F({\rm Fe\,I}) = 0.99_{0.84}^{1.12} \times 10^{-4}$ \ph, for the continuum given by bremsstrahlung emission without an absorption edge from neutral iron (see Table~\ref{t:results_noedge}).

Finally, in Section~\ref{sec:edge}, we have investigated how the presence 
of the absorption edge from neutral iron at 7.1 keV can affect the obtained constrain on the \nikel\ line flux.
For this purpose, we parametrize the absorption edge model by the normalization parameter $N_{{\rm edge}}$, defined as 
the fraction of absorbed photons along the line of sight in units of the photon flux of \fe\ fluorescent line (equations~\ref{eq:edge} and~\ref{eq:Nedge}). Repeating the fitting procedure for the given $N_{{\rm edge}}$ parameter on a grid of values from 0 to 10, 
we obtain the best agreement of the model with the observed spectrum at the parameter $N_{{\rm edge}}=5$. 
In this case, the fluorescent line fluxes are found to be
$F({\rm Ni\,I}) = 0.11^{0.01}_{0.25} \times 10^{-5}$ \ph\ and $F({\rm Fe\,I}) = 0.86_{0.72}^{1.11} \times 10^{-4}$ \ph\ 
(see Table~\ref{t:results}).

At the same time, the relative abundance of nickel in the jets of the system is determined to be equal to $Z_{{\rm Ni}}/Z = 10.1_{8.5}^{11.6} $. Proceeding from the above, we come to the conclusion that the relative abundance of nickel in the SS~433 wind should be much less than the observed abundance of nickel in the jets: for the continuum model without the absorption edge --- $Z_{{\rm Ni, wind}}/Z_{{\rm Ni, jet}}   = 0.08_{0.00}^{0.24}$ and for the continuum model with the absorption edge --- $Z_{{\rm Ni, wind}}/Z_{{\rm Ni, jet}} = 0.28^{0.62}_{0.03}$.

\section*{Acknowledgements}
The research was supported by the Russian Science Foundation (project no. 14-12-01315).

\bibliographystyle{mnras}
\bibliography{bib_en}



\appendix
\section{Method of data analysis}
\label{sec:append}
The data analysis was performed by means of the standard spectral analysis tools of the {\sc XSPEC} package \citep[version 12.9.1m,][]{Arnaud1996}.  We apply the \textit{wabs} multiplicative model for interstellar absorption to all models described in the work,  fixing the parameter of equivalent hydrogen column to $n_H = 1.2\times 10^{22}$ cm$^{-2}$ \citep{KMS2016}.
The thermal bremsstrahlung continuum was calculated by the \textit{bremss} model.
When fitting with the \texttt{lbjet} model, Doppler shifting was performed using the convolution model \textit{zashift},
which also accounts for the corresponding relativistic boosting.  The line Doppler broadening was imposed by the \textit{gsmooth} convolution model. We used the \textit{edge} model to describe the absorption edge from neutral iron.

In this work we use spectra without binning the energy channels.
In order to estimate unbaised parameters based on the $\chi^2$-statistic,
we apply the weighting function by \cite{Churazov1996}.
Besides that, we have checked that analysis of the re-binned spectra with at least 25 raw counts per bin yields results consistent within the ranges indicated by the uncertainties.

The reliability of the result obtained and the degree of degeneracy of the individual parameters of the models were determined using  the Markov Chain Monte Carlo method (MCMC). As a scheme of the Markov chain, we adopt the Metropolis-Hastings algorithm  \citep{Hastings1970}. 
The proposal distribution, from which the initial chain are generated, was set up by means of the covariance matrix at the best-fit point, obtained by the method described in Section~\ref{sec:results}. 
Such a covariance matrix was preliminarily multiplied by 
a factor of 0.2 in order to random walks better cover the model parameter space.
Therefore, it is important to discard a sufficient number of initial steps $N_{{\rm burn}}$, so that the resulting posterior parameter distributions fall in the vicinity of the maximum of a likelihood function.
We set the $N_{{\rm burn}}$ parameter in the range from $10^4$ to $5\cdot 10^4$ steps, verifying that this number is indeed sufficient for MCMC to reach a steady state distribution of $\chi^2$-statistic values.
The constraints on model parameters presented in this work (including the uncertainties on the best-fit line fluxes) correspond to the ranges covering the values of the parameters with a given posterior probability, in this paper intervals between 5\% and 95\% quantiles are used everywhere (90\% confidence level). The average value of the parameters was determined as the average over a posterior distribution. When analyzing 10 ks spectra, the final values of the model parameters, including the flux ratios of the \nikel\ and \fe\ lines,  $Z_{{\rm Ni}} / Z$,  were determined from the distributions grouped from the posterior distributions for individual 10 ks spectra.

\begin{landscape}
\begin{figure*}
\centering
\hspace*{-3cm}  
\includegraphics[width=1.3\textwidth]{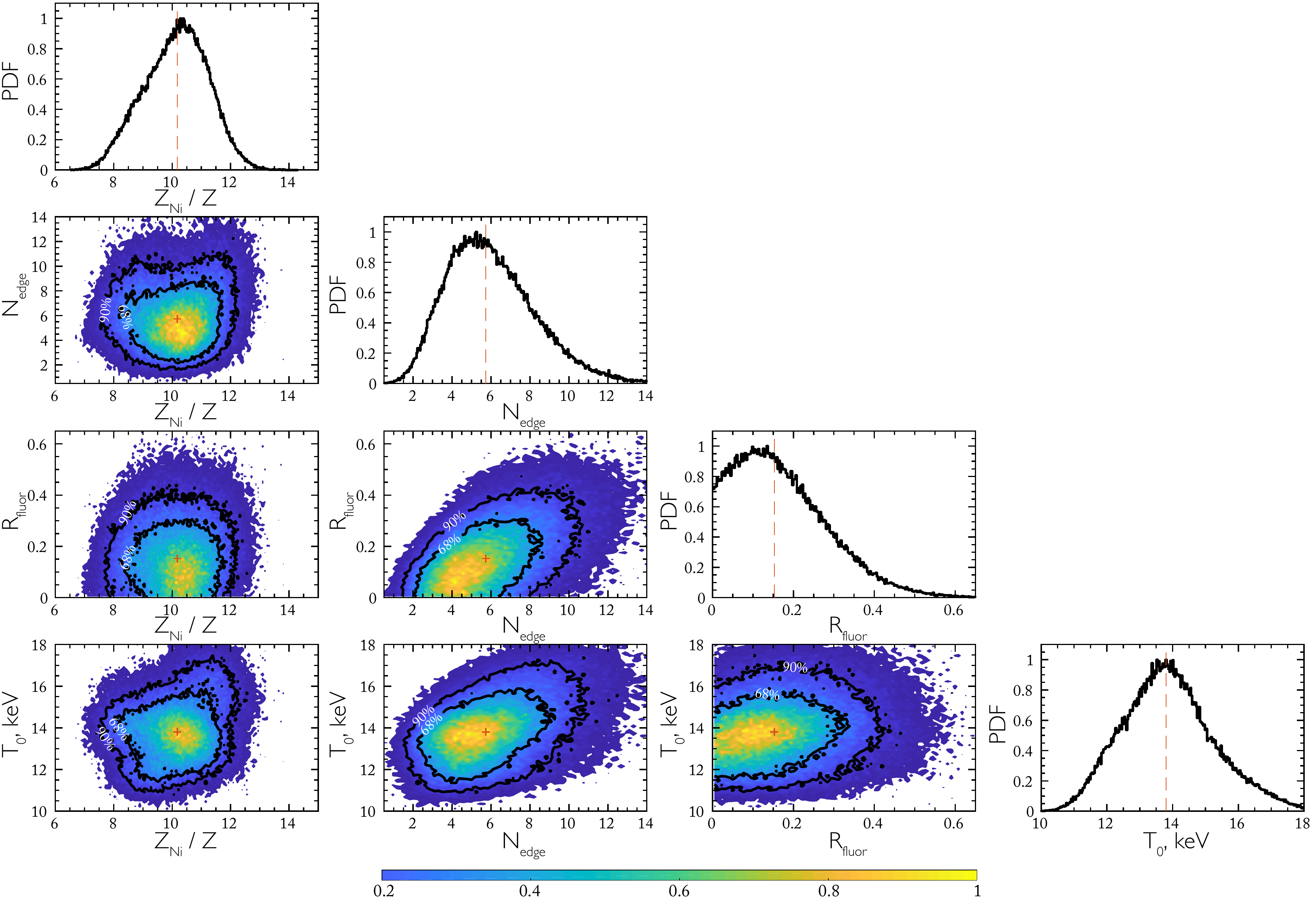} 
\caption{\small
The one- and two-dimensional marginalized posterior probability distributions of the \texttt{lbjet} model parameters obtained by fitting of the sum of the model and the continuum given by equations~\ref{eq:edge}~and~\ref{eq:Nedge} to 
 the data in the 4.3--12 keV range. The distributions are grouped from the corresponding distributions for individual 10 ks spectra. Vertical dashed lines and red crosses show the median mean values of the parameters. Black contours on the two-dimensional distributions show 90\% and 68\%  probability regions. The color scale indicates the probability density of parameters with the corresponding colour.}
\label{f:margin}
\end{figure*}

\end{landscape}

In Section~\ref{sec:edge} we discussed the effect of the neutral iron absorption edge on the obtained best-fit parameters: the flux ratio of fluorescent lines and the relative abundance of nickel in jets. 
In addition to the procedure described in that section, in which the depth of the absorption edge, expressed in units of \fe\ fluorescence line flux, was set on a grid of $N_{{\rm edge}}=0$--10, we analyzed the data using similar models, but considering $N_{{\rm edge}}$ as a free parameter. In this case, the parameters of the continuum, the bremsstrahlung temperature and normalization, were fitted simultaneously with the spectral lines in the 4.2--12 keV range. 
In Fig.~\ref{f:margin} we show the one- and two-dimensional marginalized posterior distributions of the \texttt{lbjet} model parameters: the relative abundance of nickel $Z_{{\rm Ni}}/Z$, the depth of absorption edge $N_{{\rm edge}}$, the fluorescent line flux ratio  $R_{{\rm fluor}}$ and the jet base temperature $T_0 = T_{0,b} =T_{0,r}$ (assumed equal in the fit). The final parameter distributions  were obtained by grouping distributions for individual 10 ks spectra. The model achieve   $\chi^2$/d.o.f$ = 1523/1524$ at the mean values of the parameters. Because of the high degeneracy of the $N_{{\rm edge}}$ parameter, the two-dimensional distributions have a shape elongated along this parameter. The obtained average values of the parameters are: $T_{0,b} = T_{0,r} =  13.9_{11.7}^{16.42}$ keV,  $T_{{\rm bremss}} = 27.3_{18.2}^{40.5}$ keV, $N_{{\rm edge}} = 5.7_{2.7}^{10.2}$,  $F({\rm Fe\,I\ K}_\alpha)$ = $0.8_{0.6}^{1.2} \times 10^{-4}$ \ph, $F({\rm Ni\,I\ K}_\alpha)$ = $0.139_{0.016}^{0.296} \times 10^{-4}$ \ph,  $R_{{\rm fluor}} = 0.17_{0.02}^{0.39}$, 
$Z_{{\rm Ni}}/Z =  10.1_{8.2}^{11.8}$, $Z_{{\rm Ni, wind}}/Z_{{\rm Ni, jet}} = 0.37_{0.04}^{0.85}$. 
The bremsstrahlung temperature of the continuum  increases with an increase in the absorption edge depth, therefore $T_{{\rm bremss}}$ parameter is poorly constrained.

\label{lastpage}

\clearpage
\thispagestyle{empty}
\onecolumn

\end{document}